\documentclass[twocolumn]{aastex701}
\shorttitle{Coronal outflow produces mm excess}
\shortauthors{A.M. Hankla et al.}
\submitjournal{ApJ}

\usepackage{amsmath, amssymb, mathtools}

\begin{document}

\title{An outflow from the X-ray corona as the origin of millimeter emission from radio-quiet AGN}

\author[orcid=0000-0001-9725-5509,sname='Hankla']{Amelia M. Hankla}
\altaffiliation{NASA Hubble Einstein Fellow}
\affiliation{Department of Astronomy, University of Maryland, 4296 Stadium Dr. Ste 1113, College Park, MD 20742, USA}
\email[show]{lia.hankla@gmail.com}  

\author[orcid=0000-0001-7801-0362,sname='Philippov']{Alexander Philippov}
\affiliation{Department of Physics, University of Maryland, College Park, MD, USA}
\affiliation{Joint Space-Science Institute, University of Maryland, College Park, MD, USA}
\email{sashaph@umd.edu}

\author[orcid=0000-0001-9475-5292,sname='Mbarek']{Rostom Mbarek}
\altaffiliation{Spitzer Fellow}
\affiliation{Department of Astrophysical Sciences, Princeton University, Princeton, NJ 08544, USA}
\email{rmbarek@princeton.edu}

\author[orcid=0000-0002-7962-5446,sname='Mushotzky']{Richard F. Mushotzky}
\affiliation{Department of Astronomy, University of Maryland, 4296 Stadium Dr. Ste 1113, College Park, MD 20742, USA}
\affiliation{Joint Space-Science Institute, University of Maryland, College Park, MD, USA}
\email{rmushotz@umd.edu}

\author[orcid=0000-0003-1984-189X,sname='Musoke']{G. Musoke}
\affiliation{Canadian Institute for Theoretical Astrophysics, 60 St George Street, Toronto, ON M5S 3H8, Canada}
\email{g.musoke@utoronto.ca}

\author[orcid=0000-0002-5408-3046,sname='Gro{\v{s}}elj']{Daniel Gro{\v{s}}elj}
\affiliation{Centre for mathematical Plasma Astrophysics, Department of Mathematics, KU Leuven, B-3001 Leuven, Belgium}
\email{daniel.groselj@kuleuven.be}

\author[orcid=0000-0003-4475-9345,sname='Liska']{Matthew Liska}
\affiliation{Center for Relativistic Astrophysics, Georgia Institute of Technology, Howey Physics Bldg., 837 State Street NW, Atlanta, GA 30332, USA}
\affiliation{Institute for Theory and Computation, Harvard University, 60 Garden Street, Cambridge, MA 02138, USA}
\email{mliska3@gatech.edu}

\begin{abstract}
Recent observations of radio-quiet active galactic nuclei (RQAGN) have shown the presence of millimeter emission, whose origin remains unknown, from within parsec scales of the central black hole.
We argue that the mm emission comes from a spatially extended region that is magnetically connected to the compact X-ray corona, \added{in analogy to the solar wind and corona}. 
We present an analytic model scaled to corona values in which \added{non-equipartition} electrons from multiple heights along an extended conical outflow shape the mm emission.
\added{In this model,} the 100 GHz emission originates from within $\lesssim10^4$ gravitational radii ($r_g$) of the central black hole, though the projected distance from the black hole can be as low as $50r_g$ depending on the line-of-sight.
Our model predicts a flat emission spectrum $F_{\nu}\sim{\rm const}$ and a mm-to-X-ray luminosity ratio $L_{\rm mm}/L_X\sim10^{-4}$, consistent with observations.
These quantities depend weakly on the underlying electron power-law distribution function and black hole mass.
\added{We demonstrate this model's plausibility using a general relativistic magneto-hydrodynamic (GRMHD) simulation of a thin accretion disc as a case study.}
Our model highlights the need to study continual dissipation along the outflow to connect the X-ray- and mm-emitting regions. 
\end{abstract}

\keywords{\uat{Accretion}{14} --- \uat{Astrophysical black holes}{98} --- \uat{High Energy astrophysics}{739} --- \uat{Magnetic fields}{994} --- \uat{Non-thermal radiation sources}{1119} ---  \uat{Radio quiet quasars}{1354} --- \uat{Supermassive black holes}{1663} --- \uat{Radiative processes}{2055}}



\section{Introduction} \label{sec:intro}
Active galactic nuclei (AGN) emit radiation across a broad range of the electromagnetic spectrum, from radio to X-rays and even gamma-rays.
In particular, the mere presence of X-rays in galaxies is the most reliable way to identify an AGN in the first place~\citep{kara2025}.
The presence of X-rays in the spectrum suggests a component beyond the standard thin, UV-emitting accretion disc proposed by~\citet{shakurasunyaev}.
The X-rays typically have a power-law photon index $\Gamma\sim1.5-2.5$ and a cut-off around $15-200$ keV~\citep{laha2025review}.
The canonical model for producing these X-rays assumes that a spatially compact ``corona'' hosts hot, $\sim100$ keV electrons that Compton upscatter disc photons to produce the observed hard X-ray photon spectrum and thermal cut-off~\citep{sunyaev1985, haardt1991}. 
Subsequent models suggested that the corona likely contains both thermal as well as nonthermal particles~\citep{fabian2017}, the origin of which is presently not fully understood.

The compact X-ray corona holds the key to understanding the region closest to the central black hole, where most of the accretion power is released. 
For example, radiation from the corona that illuminates the disc creates the reflection spectrum, an important part of the AGN X-ray spectrum that could also help measure black hole spin~\citep{reynolds2021, bambi2021}. 
By mediating the advection of magnetic flux, the corona could also be important for jet launching~\citep{1es} and angular momentum transport in the accretion disc~\citep{shafee2008, noble2009, penna2010}.
Because the corona is likely heated through a combination of magnetic turbulence and reconnection~\citep{galeev1979, nattila2024, groselj2024}, coronal radiation provides an opportunity to probe the dissipation processes in a strongly magnetized, collisionless plasma. 

Although most studies of the corona have focused on understanding its X-ray emission, the inferred coronal magnetic field strength suggests that the nonthermal electrons in the corona should emit synchrotron emission in the radio to mm wavelength ranges.
This low frequency emission could plausibly carry observational markers of coronal properties such as magnetization. 
Radio-quiet AGN (RQAGN) provide a testbed for detecting synchrotron emission from the vicinity of the black hole, since they lack strong jets that would dominate the radio/mm wavelengths.
RQAGN comprise approximately 90\% of all AGN~\citep{padovani2016, kellermann2016}, making them promising candidates for understanding the processes that power X-rays and other wavelengths~\citep{laha2025review, kara2025}.
More than 95\% of the hard X-ray selected AGN in the BAT AGN Spectroscopic Survey have a mm component, which is strongly correlated with the X-ray component~\citep{ricci2023, kawamuro2022}, suggesting that the two regions are physically related.
Understanding the relationship between the mm- and X-ray-emitting regions will give new insights into the physics near the black hole.

Observations at radio to mm wavelengths in radio-quiet AGN have been interpreted as synchrotron emission from electrons in the corona~\citep{laor2008, behar2015, inoue_unveiling_2014, raginski2016, behar2018}, suggesting that the synchrotron-emitting electrons are located in the same region as the thermal electrons that scatter photons to X-rays.
However, as we show in Sec.~\ref{ssec:nomm}, if coronal X-rays are produced by magnetic dissipation, the resulting optical depth to synchrotron self-absorption (SSA) is quite large.
This large optical depth requires that mm emission comes from regions further from the black hole than the $10r_g$ size of the corona as constrained by X-ray reverberation and micro-lensing studies~\citep{chartas2009, uttley2014}, where $r_g\equiv GM/c^2$.

Overcoming synchrotron self-absorption in the mm requires either expanding the X-ray emitting-region to hundreds or thousands of gravitational radii~\citep{shablovinskaya2024, palacio2025} or rethinking the geometry and definition of the corona. 
Though the exact geometry of the X-ray corona remains unknown, popular models include treating the corona as the base of a jet~\citep{markoff2005} or as hot gas ``sandwiching'' the thin disc, potentially outflowing~\citep{beloborodov1999}, or as an inner hot accretion flow~\citep{esin1997}.
In the X-ray binary (XRB) context, general relativistic magnetohydrodynamic (GRMHD) simulations with Monte Carlo radiation have shown that a hot, strongly magnetized accretion flow can provide the thermal hot electron population that produces the hard state spectrum~\citep{dexter2021}.
This hot flow could transition to a thin accretion disc at some radius~\citep{shapiro1976}, with the truncation being due to Coulomb decoupling~\citep{hankla2022b, hankla2025} or the inner disc becoming magnetically arrested (MAD)~\citep{liska2022, scepi2024}, which can cause the disc to appear observationally truncated~\citep{scepi2024b}.
Notably, these GRMHD simulations also produce subrelativistic outflows both for hot advection-dominated flows and thin Shakura-Sunyaev discs~\citep{avara2016, scepi2024, dhang2025, liska2022}, as predicted from theoretical considerations~\citep{adios, bp}.

The presence of accretion disc outflows presents a natural resolution to the synchrotron self-absorption problem.
Rather than requiring that the X-ray radiation and synchrotron emission come from the same compact ($\lesssim10r_g$) region, we propose that the synchrotron-emitting electrons lie in an expanding outflow just outside the compact corona.
In contrast to the standard notion of an accretion disc wind originating from all radii in the disc, this outflow is directly connected to the inner $10r_g$ X-ray-emitting region via magnetic fields. 
The expansion of the initially strongly magnetized plasma leads to a decrease of the magnetic field and electron density with distance from the corona. 
We find that a reasonable choice of the decreasing magnetic and density profiles, as motivated by GRMHD simulations, can explain the mm emission.

In the rest of the paper, we first demonstrate the need for a new model, showing that mm emission cannot come from the X-ray corona and proposing that the mm emission originate from an extended region that is connected to the X-ray corona via magnetic fields (Sec.~\ref{sec:motivation}).
Next, we outline an analytic model for an outflow from the compact X-ray corona with continual dissipation (Sec.~\ref{sec:model}).
We show that such an outflowing region exists in GRMHD simulations and provide a case study for the magnetic field and density profiles within the outflow (Sec.~\ref{sec:grmhd}).
We then discuss observational implications and compare to previous work (Sec.~\ref{sec:obs}).


\section{Motivation for a Multizone Outflow from the X-ray Corona}\label{sec:motivation} 
We will first argue that the X-ray corona must be magnetically-powered (Sec.~\ref{ssec:stronglymag}), and then discuss the implications for mm emission.
We then demonstrate that in a magnetically powered scenario, the mm emission is produced outside the compact ($\sim10r_g$) X-ray corona, since otherwise the mm emission would be synchrotron self-absorbed (Sec.~\ref{ssec:nomm}) or the mm-emitting nonthermal electrons (Sec.~\ref{ssec:nonthermal}) would cool too quickly to reach the required energies for mm-wavelength synchrotron emission (Sec.~\ref{ssec:cooling}).

AGN coronae have electron and proton populations with different temperatures. 
Rapid radiative cooling maintains the electrons at temperatures of $\approx10^9$ K ($\approx 100$ keV).
Meanwhile, the protons are likely close to virial temperatures at radius $r$:
\begin{align}
    T_p&\approx T_{\rm virial} = \frac25\frac{m_pc^2}{k_B}\left(\frac{r}{r_g}\right)^{-1}=4\times10^{11}~{\rm K}~\left(\frac{r}{10r_g}\right)^{-1}, \label{eq:Tvirial}\\
    \theta_p&\equiv\frac{k_BT_p}{m_pc^2}=0.04\frac{T_p}{4\times10^{11}~{\rm K}}, \label{eq:thetap}
\end{align}
where we defined the dimensionless proton temperature $\theta_p$.

When the corona is sufficiently magnetized, the protons will not equilibrate with the electrons through Coulomb collisions because the thermalization time is too long relative to the turbulence dissipation time, given by the Alfv\'en crossing time $t_{\rm A}=R_c/v_A=\sigma_i^{-1/2}R_c/c$ of the compact corona.
Here $v_A=B/(4\pi\rho)^{1/2}=\sigma_i^{1/2}c$ is the non-relativistic Alfv\'en speed and $\sigma_i=B^2/(4\pi m_p nc^2)$ is the plasma ion magnetization.
Taking $t_{\rm therm}\equiv1/\nu_{\rm therm}$ as the electron-ion thermalization time~\citep{nrl}, we have:
\begin{align}
    t_{\rm therm}&=6\times10^3~{\rm s}~ \left(\frac{\tau_0}{1}\right)^{-1}\left(\frac{R_c}{10r_g}\right)\left(\frac{M}{10^7M_\odot}\right)\left(\frac{\theta_e}{0.2}+\frac{\theta_p}{0.04}\right)^{3/2},\\
    \frac{t_{\rm therm}}{t_{\rm A}}&=5~ \left(\frac{\tau_0}{1}\right)^{-1}\left(\frac{\theta_e}{0.2}+\frac{\theta_p}{0.04}\right)^{3/2}\left(\frac{\sigma_i}{0.1}\right)^{1/2},
\end{align}
where $\theta_e\equiv k_BT_e/(m_ec^2)$ in analogy with equation~\eqref{eq:thetap} and we have set the number density via the optical depth to Thomson scattering: 
\begin{equation}
    n=\frac{\tau_0}{\sigma_T R_c}=10^{11}~{\rm cm^{-3}}\left(\frac{\tau_0}{1}\right)\left(\frac{R_c}{10r_g}\right)^{-1}\left(\frac{M}{10^7M_\odot}\right)^{-1}. \label{eq:n0tau}
\end{equation}
Our model will always scale densities relative to the density in the corona according to equation~\eqref{eq:n0tau}.
Notably, the ratio $t_{\rm therm}/t_{\rm A}$ is independent of black hole mass, meaning that the compact corona will have two distinct temperatures for electrons and ions provided the corona is sufficiently magnetized ($\sigma_i\gtrsim0.1$).

\subsection{The X-ray Corona is Strongly Magnetized} \label{ssec:stronglymag}
The idea that the corona must be strongly magnetized is not new. 
Theoretical estimates have long demonstrated the need for strong coronal magnetic fields from an energetics perspective~\citep{merloni2001, beloborodov2017}.
Launching relativistic jets and winds also likely requires strong magnetic fields~\citep{bz, bp, tchekhovskoy2011}, as does producing neutrinos while absorbing gamma rays from the compact corona~\citep{murase2022,mbarek2024}.
Observational evidence for these values of magnetic fields exists in the literature measuring jet magnetic fields at large scales. 
The magnetic field at parsec and 0.1-parsec scales in an AGN jet was found to be on the order of 10G~\citep{martividal2019} and 0.1G~\citep{osullivan2009}, respectively.
Assuming the magnetic field strength falls inversely with distance from the black hole and extrapolating back to horizon scales yields magnetic fields $\gtrsim10^5$ G.

Magnetic energy provides the most plausible internal energy source for the X-ray corona.
By drawing an analogy to the solar corona and solar wind, we envision several mechanisms operating in the corona-outflow system that convert stored magnetic energy into plasma energy.
Reconnection models typically treat the corona as loops of magnetic field above the accretion disc~\citep{galeev1979, uzdensky2008}.
These loops can then reconnect when their footpoints switch, as in interchange reconnection in the solar corona~\citep{bale2023, drake2025}.
Turbulence models, inspired by the wave-heating models of the solar corona~\citep{wentzel1974, bourouaine2024}, suppose that Alfv\'en waves propagate along open field lines and dissipate via an imbalanced Alfv\'enic cascade~\citep{chandran2018, squire2022}.

Arguing that the magnetic energy density must be in equipartition with the electron pressure and using equation~\eqref{eq:n0tau} yields estimates for the magnetic field of:
\begin{align}
    B_0&\ge B_{\rm eq}^e\\
    &=700~{\rm G}~\left(\frac{R_c}{10r_g}\right)^{-1/2}\left(\frac{M}{10^7M_\odot}\right)^{-1/2}\left(\frac{T_e}{100~{\rm keV}}\right)^{1/2}. \label{eq:Beq}
\end{align}
However, equipartition with electrons does not hold in the strongly magnetized corona.
The energy dissipated from magnetic turbulence/reconnection into electron heat passes almost instantaneously into photons~\citep{werner2019, groselj2024} because of electrons' short cooling times (see Sec.~\ref{ssec:cooling}).
Instead assuming equipartition with protons at the virial temperatures such that $3/5 GM/R_c=3/2 k_BT_p/m_p$ yields higher magnetic field estimates:
\begin{equation}
    B_0\approx B_{\rm eq}^p=2\times 10^4~{\rm G}~\left(\frac{R_c}{10r_g}\right)^{-1}\left(\frac{M}{10^7M_\odot}\right)^{-1/2}\left(\frac{\tau_0}1\right)^{1/2}.
\end{equation}

This estimate of equipartition with protons more closely matches the X-ray power from the conversion of magnetic energy into X-rays with an efficiency $\eta_X<1$: 
\begin{align}
  L_X&= \eta_X\frac{B_0^2}{8\pi}4\pi R_c^2c \label{eq:LX}\\ 
  &=3\times10^{43}~{\rm erg/s}\left(\frac{\eta_X}{0.1}\right)\left(\frac{B_0}{10^4~{\rm G}}\right)^2\left(\frac{M}{10^7M_\odot}\right)\left(\frac{\tau_0}1\right), 
\end{align}
where typical X-ray (14 - 150keV) luminosities of radio-quiet AGN range from $10^{41}{\rm erg/s}$ to $10^{44}{\rm erg/s}$~\citep{ricci2023}. 
\added{Taking $\eta_X<1$ accounts for the fact that some magnetic energy converts into proton heating rather than X-rays.
This lower bound also assumes that electrons cool on timescales faster than $R_c/c$, which we will justify in Sec.~\ref{ssec:cooling}.}
The collisionless reconnection rate provides an upper bound on this efficiency factor of 0.1~\citep{werner2018}.

Requiring $\eta_X<1$ leads to a {\it lower bound} on the magnetic field of:
\begin{equation}
    B_0\geq10^3~{\rm G}\left( \frac{L_X}{10^{43}~{\rm erg/s}} \right)^{1/2}\left( \frac{R_c}{10r_g} \right)^{-1}\left( \frac{M}{10^7M_\odot} \right)^{-1}, \label{eq:Blower}
\end{equation}


\subsection{Synchrotron Self-Absorption Prevents the X-ray Corona From Producing Observable Millimeter Emission} \label{ssec:nomm}
The strong magnetic field given by equation~\eqref{eq:Blower} means that any synchrotron emission emitted from within the $10r_g$ X-ray-emitting region will be strongly self-absorbed.
Below the synchrotron turnover frequency $\nu_t$, the flux from a population of nonthermal electrons in a sphere with uniform magnetic field and density drops as $\nu^{5/2}$~\citep{rybickilightman}.
The synchrotron turnover frequency within the strongly magnetized corona is large enough that this $\nu^{5/2}$ drop wipes out any observable emission at 100 GHz.
Explicitly, the synchrotron turnover frequency is determined by setting the optical depth to SSA equal to 1: $\tau_{\rm SSA}=1=\alpha_\nu R_c$ (see Appendix~\ref{app:sourcefunc}).
To compare with previous models of coronal synchrotron emission~\citep{laor2008, inoue2018}, we assume a nonthermal particle distribution with a power-law index $p=2$ that comprises 1\% of the total \added{electron number} density. 
Setting the total number density through $\tau_0=1$ (equation~\eqref{eq:n0tau}) yields:
\begin{equation}
    \nu_t=10^{4}~{\rm GHz}~\left(\frac{B}{10^3~{\rm G}}\right)^{2/3}\left(\frac{R_c}{10r_g}\right)^{1/3} \label{eq:nutsphere}
\end{equation}
where $B_0$ and $n_0$ are the value of the uniform magnetic field and number density, respectively.
From equation~\eqref{eq:nutsphere}, we see that the synchrotron turnover frequency within the compact corona reaches values of $10^4$ GHz, meaning that the flux at 100 GHz will have dropped by a factor of $(10^4~{\rm GHz}/10^2~{\rm GHz})^{5/2}\sim10^5$ from its peak value. 

\begin{figure}
    \centering
    \includegraphics[width=0.95\linewidth]{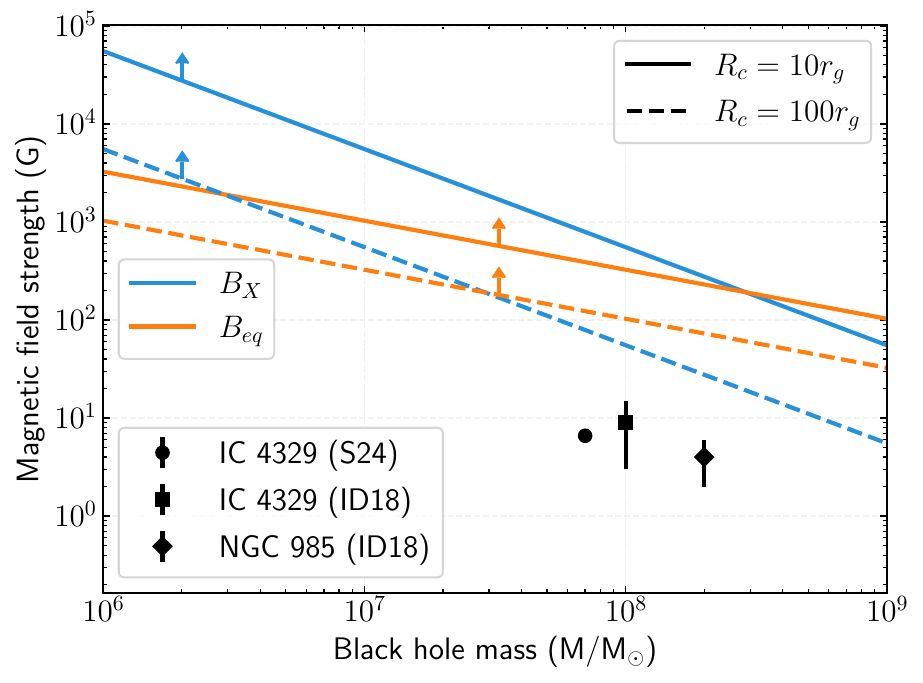}
    \caption{Lower bounds on the magnetic field at $10r_g$ for a strongly magnetized corona with size $R_c$. The constraint $B_X$ comes from converting magnetic energy into X-ray luminosity (equation~\eqref{eq:LX}), while $B_{\rm eq}^e$ comes from the requirement that electrons are colder than equipartition magnetic fields because they radiate (equation~\eqref{eq:Beq}). For comparison, recent observational fits to magnetic field are shown in black markers. Here, we abbreviate~\citet{inoue2018} as ID18 and~\citet{shablovinskaya2024} as S24. Arrows indicate that these lines yield lower bounds. This figure assumes $L_X=10^{43}~{\rm erg/s}$~\citep{shablovinskaya2024}, $\tau_0=1$, and $T_e=10^9~{\rm K}$.}
    \label{fig:Bbounds}
\end{figure}

A number of proposed models invert a variation of equation~\eqref{eq:nutsphere} to estimate the coronal magnetic field strength.
These models assume that the peak mm emission, which occurs around 100 GHz, represents the synchrotron turnover frequency of the uniform, spherical, synchrotron-emitting plasma.
equation~\eqref{eq:nutsphere} then yields magnetic fields on the order of 1G - 10G and emitting region sizes of $10^2$-$10^3r_g$~\citep{behar2015, inoue2018, petrucci2023, shablovinskaya2024, palacio2025}.
These large coronal sizes disagree with those inferred from X-ray observations, with tentative simultaneous variability suggesting that the mm- and X-ray-emitting regions are connected but separate~\citep{petrucci2023}.
The low values of magnetic field violate the bounds set by conversion of magnetic energy into X-rays (equation~\eqref{eq:LX}) and by equipartition (equation~\eqref{eq:Beq}), as demonstrated by Figure~\ref{fig:Bbounds}.

The low values of magnetic field suggest two possibilities: first, that equation~\eqref{eq:nutsphere} needs to be revised, or second, that the corona is not magnetically-powered. 
Having already motivated why the X-ray corona must be magnetically-powered (see also Sec.~\ref{ssec:cooling}), the majority of this work focuses on revising equation~\eqref{eq:nutsphere} to maintain the physically-motivated assumption of a magnetically-powered corona. 
To do so, we drop equation~\eqref{eq:nutsphere}'s assumption that the mm-emitting electrons are co-located with the X-ray-scattering electrons. 
Instead, we consider an extended region outside the compact X-ray corona that has a decreasing magnetic field and density profile. 
Further from the black hole, the turnover frequency drops, allowing the emission to remain optically thin at lower frequencies.

\subsection{Millimeter Emission Requires Nonthermal Electrons} \label{ssec:nonthermal}
The characteristic frequency of synchrotron emission $\nu_s$ of an electron with Lorentz factor $\gamma$ in a magnetic field $B$~\citep{rybickilightman} shows that producing 100 GHz synchrotron emission requires electrons that have Lorentz factors $\gamma\approx 100$, i.e. nonthermal electrons:
\begin{align}
    \nu_s=\gamma^2\frac{eB}{2\pi m_ec}=100~{\rm GHz}~\left(\frac{\gamma}{100}\right)^2\left(\frac{B}{4~{\rm G}}\right). \label{eq:nus}
\end{align}
In our model, the outflow's strong field strengths initially have the compact X-ray corona values (equation~\eqref{eq:Blower}), and decay to $\approx 4~{\rm G}$ at a certain distance, thereby producing 100 GHz emission at distances $r>10r_g$ from the black hole. 

However, recent kinetic plasma physics simulations of particle acceleration processes reveal that magnetic fields alone do not determine the presence of nonthermal electrons or their highest achievable Lorentz factor.
Instead, the ratio of magnetic energy density to particle rest-mass energy $\sigma$ determines properties of the nonthermal electron population.
Relating the electron magnetization $\sigma_e$ to the magnetic field shows that: 
\begin{equation}
    \sigma_{e}\equiv \frac{B^2}{4\pi n_e m_ec^2}=100\left(\frac{B}{10^4~{\rm G}}\right)^2\left(\frac{R_c}{10r_g}\right)\left(\frac{M}{10^7M_\odot}\right)\left(\frac{\tau_0}1\right)^{-1}, \label{eq:sigmaB}
\end{equation}
where the number density $n_e$ was set using the Thomson scattering optical depth via equation~\eqref{eq:n0tau}.
\added{Equation~\eqref{eq:sigmaB} shows that the average magnetic energy per particle is approximately $\sigma_e$.}
\added{Particle injection to this mean energy, i.e. up to $\gamma\approx\sigma_e$, occurs extremely rapidly in 3D numerical particle-in-cell (PIC) simulations~\citep{sironi2022, french2023}.
Beyond this mean energy, the formation of a power-law tail at Lorentz factors $\gamma\gg\sigma_e$ depends sensitively on the guide field and cooling. 
Simulations of a single current sheet undergoing relativistic reconnection with no guide field show a maximum Lorentz factor extending far beyond $\sigma_e$~\citep{zhang2021}, even in the presence of strong cooling~\citep{chernoglazov2023}.
Whether nonthermal particles beyond $\gamma\gg\sigma_e$ exist in simulations of radiative turbulence, which are more applicable to the wind scenario, remains an active field of research, with contradictory answers in the literature~\citep{zhdankin2020, lemoine2025}. 
In principle, slow diffusive acceleration from scattering off magnetic mirrors in turbulence could produce larger high-energy tails~\citep{fermi1949, zhdankin2018, vega2024}.
However, in practice, fast cooling times may limit these acceleration mechanisms (see Sec.~\ref{ssec:cooling}).
Because the presence of nonthermal particles depends more on the mean energy than the maximum achievable Lorentz factor, we will conservatively require $\sigma_e\gg1$ to produce nonthermal electrons with $\gamma\sim100$.
}
Notably, models of the corona with weak magnetic fields have values of $\sigma_e\lesssim1$~\citep{palacio2025, inoue2018}. 
Given our current understanding of particle acceleration in magnetized plasmas, such low magnetizations present challenges for reaching the required $\gamma\sim100$ energy range. 

Multiwavelength and multimessenger observations suggest the presence of nonthermal electrons in X-ray binary (XRB) and AGN coronae. 
X-ray binary systems such as Cygnus X-1  demonstrate a high-energy tail above 10 keV in the soft spectral state~\citep{mcconnell2002}, and an MeV tail in the hard spectral state~\citep{mcconnell2000}.
The high-energy emission tail is difficult to produce with an exclusively thermal electron population, motivating the need for a non-thermal population, perhaps from the collisionless plasma within the innermost stable circular orbit of the black hole~\citep{hankla2022b}.
It stands to reason that this high-energy emission tail also exists in AGN, but is too faint for observations to detect (see~\citet{ajello2009, baity1984} for observational constraints). 
In AGN, the presence of nonthermal electrons has been invoked to explain lower-than-expected coronal temperatures due to a pair thermostat~\citep{fabian2017}.
In addition, recent observations have suggested that neutrinos could originate from AGN coronae, whose radiation-rich environment absorbs the gamma-rays that would otherwise be expected from neutrino production~\citep{murase2020}. 
For neutrinos to come from the corona, the corona would have to not only accelerate electrons to nonthermal energies, but also protons~\citep[e.g.,][]{mbarek2024, fiorillo2024}.

\subsection{Fast Cooling Reduces Nonthermal Particle Acceleration in the Compact Corona} \label{ssec:cooling}
Existing particle-in-cell (PIC) simulations \added{show} only moderate amounts of nonthermal electrons in the compact corona itself due to fast radiative cooling~\citep{groselj2024, nattila2024}. 
The radiative and magnetic compactness parameters quantify the impact of cooling on nonthermal particle acceleration by comparing the ratio between the light-crossing time and the inverse Compton and synchrotron cooling time, respectively. 
Defining the cooling time of an electron with Lorentz factor $\gamma$ and velocity $\beta\to c$ due to inverse Compton scattering as $t_{\rm IC}=E/(dE/dt)$ where the cooling rate is given by $dE/dt=(4/3) \sigma_T c\gamma^2 U_{\rm ph}$~\citep{coppi1999}, we have
\begin{equation}
    t_{\rm IC}\equiv\frac{E}{dE/dt}=\frac34\frac{m_ec}{\sigma_T \gamma U_{\rm ph}} \label{eq:tIC}
\end{equation}
where $U_{\rm ph}$ is the photon energy density. 
For a spherical source of radius $R$ with luminosity $L_{\rm ph}=4\pi R^2c U_{\rm ph}/3$, the dimensionless luminosity gives the radiative compactness: $\ell=L_{\rm ph}\sigma_T/(Rm_ec^3)$. 
Combining with equation~\eqref{eq:tIC} and setting the light-crossing time $t_{\rm LC}=R/c$ shows that the radiative compactness measures the ratio of light-crossing time to Compton cooling time for an electron with a given Lorentz factor $\gamma$:
\begin{align}
    \frac{t_{\rm LC}}{t_{\rm IC}}&=\gamma\ell, \label{eq:gammaell}\\
    \ell&=6~\left(\frac{L_{\rm ph}}{10^{43}~{\rm erg/s}}\right)\left(\frac{M}{10^7M_\odot}\right)^{-1}\left(\frac{z}{10r_g}\right)^{-1}
\end{align}
where the second line assumes the radius increases linearly with $z$ and starts with a width of $10r_g$ at $z=10r_g$, i.e. $R=10r_g(z/10r_g)$.

Because the synchrotron and inverse Compton cooling times are related through the ratio of magnetic and photon energy densities $t_{\rm sync}/t_{\rm IC}=U_B/U_{\rm ph}$, the magnetic compactness gives the ratio between light-crossing time and synchrotron cooling time: $\gamma\ell_B=t_{\rm LC}/t_{\rm sync}$.
The magnetic compactness directly relates to the optical depth to Thomson scattering (equation~\eqref{eq:n0tau}) and the electron magnetization $\sigma_e=B^2/(4\pi m_e nc^2)$ as $\ell_B\sim2\sigma_e\tau_0\to2\sigma_e$ in the compact corona. 
For stronger magnetic fields, the magnetic compactness increases and therefore higher Lorentz factor electrons will cool before escaping the corona.
Assuming a magnetic field that decays inversely with distance $z$ from the black hole, the magnetic compactness as a function of $z$ reads:
\begin{equation}
    \ell_B=3\left(\frac{M}{10^7M_\odot}\right)\left(\frac{B_0}{10^4~{\rm G}}\right)^2\left(\frac{z}{100r_g}\right)^{-1} \label{eq:ellB}
\end{equation}
where $B_0$ is the magnetic field at $10r_g$ bounded by equation~\eqref{eq:Blower} and we assume that the radius of the outflow increases linearly with $z$.
This equation shows that all electrons will be strongly cooled until they reach distances of $\approx100r_g$ from the black hole, and electrons with $\gamma=100$ will be strongly cooled until a radius of $\approx10^4r_g$.
Close to the black hole, \added{synchrotron photons will be absorbed by the same population of electrons that emitted them, thereby thermalizing the electron population below a certain Lorentz factor, typically $\lesssim10$~\citep{ghisellini1998,malzac2009, beloborodov2017}}, so most of the overall cooling is done by inverse Compton scattering\added{\footnote{The population of seed photons that scatter off the hot electrons depends on the coronal model, and we do not specify it here.}}.

Observationally, the radiative compactness spans values of 1 to 100~\citep{fabian2015, hinkle2021}.
Relating the radiative and magnetic compactness by assuming, similar to equation~\eqref{eq:Blower}, that the magnetic energy converts into X-rays yields $\ell\sim\eta_X\ell_B\sim0.1\ell_B$.
We thus focus on initial values of magnetic compactness and therefore $\sigma_e$ between $10$ and $10^3$.
For values of $\sigma_e\gtrsim100$, the $\gamma=100$ Lorentz factor electrons needed to produce 100 GHz emission (equation~\eqref{eq:nus}) are in the strong cooling regime, losing most of their energy to radiation before escaping the corona.
We therefore require continued dissipation along the length of the outflow in order to continue accelerating particles to high energies that then cool via synchrotron emission.


\section{Analytic model for a millimeter-emitting Region Connected to the X-ray Corona} \label{sec:model}
This section presents an analytic model for the extended outflow from the X-ray corona motivated by the arguments given above (Sec.~\ref{sec:motivation}). 
The strongly magnetized and expanding outflow, driven by dissipation in the compact corona, extends beyond the synchrotron self-absorbed compact corona and remains magnetized enough to ultimately produce mm emission. 
While the properties of the outflow itself are motivated by GRMHD simulations (see Sec.~\ref{sec:grmhd}), the radiative properties follow the principles for flat radio-spectrum relativistic jets outlined in~\citet{bk}. 

\subsection{Example Outflow Properties} \label{ssec:properties}
\begin{figure}
  \centering
  \includegraphics[width=0.45\textwidth]{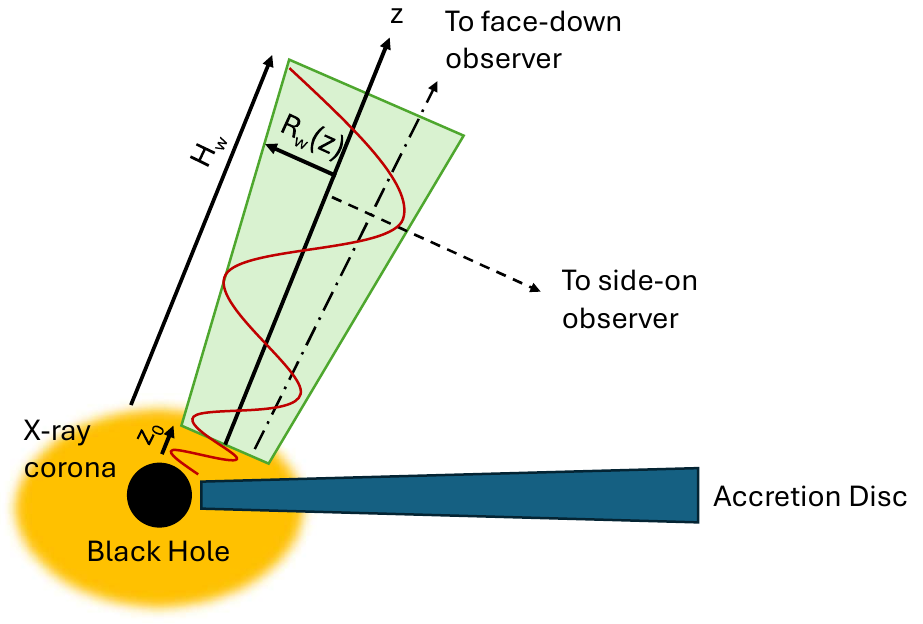}
  \caption{Diagram of the outflow set-up. The coordinate $z$ runs along the length of the outflow, while $r$ runs perpendicular to the outflow axis. The X-ray corona, shown in orange, is located somewhere within $10r_g$ of the black hole. The extended mm-emitting region, shown in green, connects to the corona via magnetic field lines (red).}
  \label{fig:schematic}
\end{figure}
As a concrete example of the expanding outflow, we parameterize the outflow properties as a function of height $z$ above the central black hole (or accretion disc midplane), as illustrated in Figure~\ref{fig:schematic}.
We treat most outflow quantities as a power law, with values at
the outflow base $z_0$ denoted with the subscript $0$.
In particular, the magnetic field perpendicular to the jet axis $B_\perp$ and electron number density $n_e$ vary as
\begin{align}
  B_\perp(z)&=B_0\left(\frac{z}{z_0}\right)^{-b} \label{eq:modelB}\\
  n_e(z)&=n_0\left(\frac{z}{z_0}\right)^{-a}. \label{eq:ne}
\end{align}
We will take the values $b=1$ and $a=2$ extracted from \added{our fiducial GRMHD simulation} (Sec.~\ref{sec:grmhd}) as an example unless otherwise noted.
The case of a constant magnetic field in a uniform density sphere can be recovered by setting $a=b=0$.
The magnetic field at the outflow base $B_0$ is set using equation~\eqref{eq:Blower}.
We estimate the electron number density at the outflow base $n_0$ by setting the optical depth to Thomson scattering $\tau_0$ over the size of the corona $R_c$ equal to 1 (equation~\eqref{eq:n0tau}):
\begin{align}
    n_0&\approx10^{11}~{\rm cm^{-3}}\left( \frac{R_c}{10r_g} \right)^{-1}\left( \frac{M}{10^7M_\odot} \right)^{-1}.
\end{align}
The profile of the plasma electron magnetization $\sigma_e$ along the outflow is given by:
\begin{align}
  \sigma_e(z)&=\frac{B^2}{4\pi n_e m_ec^2}=\sigma_{e0}\left(
  \frac{z}{z_0}\right)^{-2b+a}\to \sigma_{e0} ,\label{eq:sigmae}
\end{align}
\added{where the right arrow indicates the default values $b=1$ and $a=2$.} 
Note that for these default choices, the electron magnetization is constant along the outflow.
This choice is again motivated by the GRMHD simulations outlined in Sec.~\ref{sec:grmhd}.

The outflow must satisfy the equations of mass
conservation and conservation of magnetic flux.
These equations read:
\begin{align}
  \rho(z) R_w^2(z) v_z(z) &= \dot M_w={\rm constant}\label{eq:masscons}\\
  R_w^2(z)B_\parallel(z) &= {\rm constant} \label{eq:fluxcons}
\end{align}
From equation~\eqref{eq:fluxcons}, the width of the outflow $R_w(z)$ is determined by the
parallel magnetic field profile $B_\parallel$.
\begin{equation}
  R_w(z)=R_{w0}\left(\frac{z}{z_0}\right)^w\to R_{w0}\left(\frac{z}{z_0}\right), \label{eq:Rwz}
\end{equation}
where we set $R_{w0}=10r_g$ unless otherwise noted.
We typically set $B_\parallel\propto (z/z_0)^{-2}$, yielding $w=1$ and thus giving the outflow a conical shape. 
Similarly, the velocity profile of the outflow is constrained by equation~\eqref{eq:masscons}:
\begin{equation}
  v_z(z)=v_{z0}\left( \frac{z}{z_0} \right)^{a-b}\to v_{z0}\left( \frac{z}{z_0} \right).\label{eq:vz}
\end{equation}
Because the outflow speed is sub-relativistic, we do not include beaming effects that would come into play for a relativistic outflow speed.
As such, the exact outflow speed is unimportant to the model. 


For simplicity, we fix several geometric properties of the outflow.
In particular, we treat the outflow at each height as a square, with the distance from the outflow axis denoted by the coordinate $r$.
This radial coordinate runs from $-R_w$ to $R_w$, where $R_w$ in general can depend on the height $z$, or become the radius of a cylindrical or azimuthally-symmetric outflow. 
We scale $z_0$ and $R_{w0}$ to the radius of the corona $R_c$, which is set to $10r_g$ unless otherwise noted.
At some point along the outflow, nonthermal particle acceleration will become inefficient if the electron magnetization $\sigma_e$ drops below 1.
In this case, the final outflow height $H_w=H_w(\sigma_e=1)$.
For our canonical choice of parameters, $\sigma_e$ is constant along the outflow and we therefore set $H_w/z_0=10^7$.
We set a distance to the observer of $D=10~{\rm Mpc}$ unless otherwise noted.

We consider a hybrid thermal-nonthermal distribution of electrons, where the thermal electrons contain most of the density.
How many nonthermal electrons accelerate out of the thermal bulk depends on the energy dissipation mechanism, be it turbulence or magnetic reconnection.
We define the nonthermal number fraction as:
\begin{equation}
    \eta=\frac{n_{\rm PL}}{n_e}=\frac{n_{\rm PL}}{n_{\rm PL} + n_{\rm MJ}} \label{eq:etadef}
\end{equation}
and fix $\eta(z)=\eta_0=0.1$ unless otherwise noted.
Here, $n_{\rm PL}$ is the number density of the electrons in the power-law, while $n_{\rm MJ}$ is the number density of the thermal, Maxwell-J\"uttner electrons that dominate the overall number density.

The nonthermal distribution function extends from Lorentz factors $\gamma_1$ to $\gamma_2$ as:
\begin{equation}
\begin{array}{r r}
    N_{\rm PL}(\gamma,z)d\gamma=C\gamma^{-p}d\gamma, & \gamma_1\leq\gamma\leq\gamma_2, \label{eq:nPL}
\end{array}
\end{equation}
where $\gamma_1$, $\gamma_2$, and $p$ can depend on e.g. the plasma magnetization.
The normalization $C$ is set by the values of $n_0$ and $\eta_0$, and $z$-dependent quantities such as $\gamma_1$, $\gamma_2$, and $p$.
Requiring that
\begin{equation}
    n_{\rm PL}(z)=\int_1^\infty N_{\rm PL}(\gamma,z)d\gamma
\end{equation}
and combining with Eqns.~\ref{eq:ne} and~\ref{eq:etadef} yields 
\begin{align}
    C(z)&=\frac{p(z)-1}{\gamma_1(z)^{1-p(z)}-\gamma_2(z)^{1-p(z)}}\eta(z)n_e(z)\\
    &= f_4(p)\eta_0n_0\left(\frac{z}{z_0}\right)^{-a}\to f_4(p)\eta_0n_0\left(\frac{z}{z_0}\right)^{-2},\label{eq:Cz}
\end{align}
where $f_4(p)=(p-1)/(\gamma_1^{1-p}-\gamma_2^{1-p})\approx(p-1)$.

In this work, we consider a fixed nonthermal electron distribution function unless otherwise noted.
In principle both the high-energy cut-off $\gamma_2$ and the power-law index $p$ should depend on parameters such as the magnetization and acceleration mechanism, though as we will show in later sections, $p$ does not strongly affect the radiative properties of the outflow and we therefore treat $p$ as a constant.
Here, we set $\gamma_1=2$ and $\gamma_2=100$ in accordance with equation~\eqref{eq:nus}.

Within this framework ($a=2$, $b=1$), the Lorentz factor $\gamma_c$ of the electrons emitting $\nu_c=$100 GHz synchrotron emission is given by (equation~\eqref{eq:nus}, assuming $\tau_0=1$ at the corona):
\begin{equation}
\gamma_c=60~\left(\frac{\nu_c}{100~{\rm GHz}}\right)^{1/2}\left(\frac{\sigma_{e0}}{100}\right)^{1/2}\left(\frac{M}{10^7M_\odot}\right)^{1/4}\left(\frac{z}{10^3z_0}\right)^{1/2}.
\end{equation}
Near the base of the outflow, electrons experience strong radiative cooling, which then gradually weakens with increasing distance from the corona (equation~\eqref{eq:ellB}).
The distance $z_f$ where an electron with a given Lorentz factor stops being strongly cooled is \added{where $\ell_B(\gamma, z_f)=1$, i.e.}:
\begin{align}
    \frac{z_f}{z_0}&=6.5\times10^3\left(\frac{\gamma}{100}\right)\left(\frac{\sigma_{e0}}{100}\right)\left(\frac{\tau_0}{1}\right). \label{eq:zf}
\end{align}
Note that the combination of $\sigma_{e0}$ and $\tau_c$ means that equation~\eqref{eq:zf} is density-independent.
Therefore, 100 GHz-emitting electrons are strongly cooled until a distance $z_f(\gamma_c)$, where
\begin{equation}
    \frac{z_f(\gamma_c)}{z_0}=1.5\times10^3~\left(\frac{\sigma_{e0}}{100}\right)\left(\frac{M}{10^7M_\odot}\right)^{-1/2}\left(\frac{\nu_c}{100~{\rm GHz}}\right).
\end{equation}
Because they cool before they reach distances of $\sim10^4r_g$, the 100 GHz-emitting electrons cannot be accelerated in the compact corona and then advected away from the black hole to regions of lower magnetic field.
We therefore implicitly assume that the power law of electrons is injected at each height, i.e. the outflow must continually accelerate particles to Lorentz factors of $\gamma_c$.


\subsection{Radiative Transfer Through the Outflow} \label{ssec:rtoutflow}
\begin{figure}
    \centering
    \includegraphics[width=0.5\textwidth]{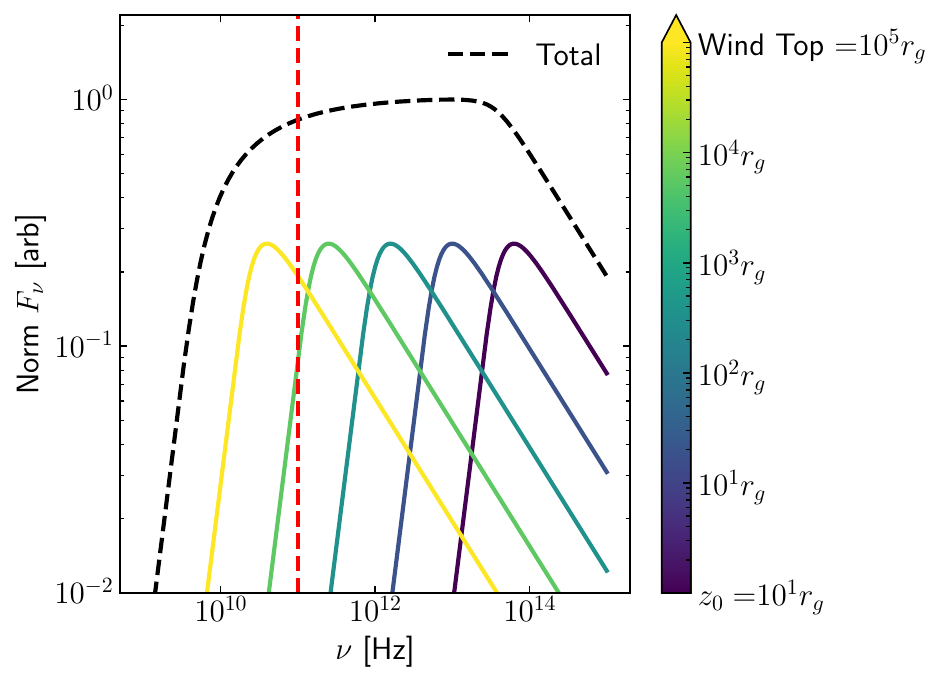}
    \caption{As a result of the inhomogeneity in the outflow, a flat spectrum develops between the turnover frequency at the outflow base and the turnover frequency at the outflow top. The specific intensity at select heights is shown by the colored lines. The outflow base has a turnover frequency of around $10^{14}$ Hz (purple peak). Above this peak, the total spectrum (dashed line) drops off as $\nu^{-(p-1)/2}=\nu^{-1/2}$. Below this peak lies the flatter, intermediate regime wherein the peaks from each outflow height sum together. Below the turnover frequency of the outflow top (yellow peak, $\approx 10$ GHz), the spectrum demonstrates the familiar $\nu^{5/2}$ dependence. 
    The flux has been normalized to its maximum value. 
    The dotted red vertical line shows $\nu=100$ GHz.
    Above $\approx300$ GHz, the flat power-law is likely obscured by dust. }
    \label{fig:sum-peaks}
\end{figure}
We calculate the radiative properties of the outflow in two limiting cases: i) the observer views the outflow perpendicular to its axis (edge-on), and ii) the observer views the outflow parallel to its axis (face-down).
In this section, we present the edge-on scenario, with details given in Appendix~\ref{app:sideon}.
The face-on case is outlined in Appendix~\ref{app:facedown}.
All outflow properties are the same except for the viewing angle.

The differential flux contributed from each height of the outflow is a classic
synchrotron self-absorption spectrum, comprising a peak at the synchrotron
turnover frequency $\nu_t(z)$, a power-law that goes as $\nu^{5/2}$ for
$\nu<\nu_t(z)$, and a power-law that goes as $\nu^{-(p-1)/2}$ for
$\nu>\nu_t(z)$.
The turnover frequency is given by
\begin{align}
\nu_t(z)&=\left[2R_w(z)\alpha_{\nu0}(z/z_0)^{-k_\alpha}\right]^{2/(p+4)}\\
&=\nu_{t0}\left(\frac{z}{z_0}\right)^{-k_\nu}\to\nu_{t0}\left(\frac{z}{z_0}\right)^{-1}. \label{eq:nut_z_scaling}
\end{align}
where $p$, $B_\perp$, $\eta$, $R_w$, and $n$ depend on $z$ as discussed in Sec.~\ref{ssec:properties}.
The last line defines $k_\nu\equiv 2(k_\alpha-w)/(p+4)\to (p+6-2w)/(p+4)$ when assuming $a=2$, $b=1$, and $d=0$.
Notably, for the conical outflow choice of $w=1$, $k_\nu\to1$ is independent of $p$; thus the dropoff in turnover frequency with height is independent of the underlying electron distribution function.
This choice of density and magnetic field drop-off shifts the peaks of larger outflow heights to lower frequencies.
The turnover frequency at the outflow base $\nu_{t0}=\nu_t(z_0)$ is given by:
\begin{align}
    \nu_{t0}&\sim\left(\frac{\sigma_{e0}}{M}\right)^{\frac{(p+2)}{2(p+4)}}\\
    &\stackrel{\mathclap{\mbox{p=2}}}{=}9.6\times 10^{13}~{\rm Hz}\left(\frac{\sigma_{e0}/100}{M/10^7M_\odot}\right)^{1/3}\\
    &\stackrel{\mathclap{\mbox{p=3}}}{=} 5.1\times10^{13}~{\rm Hz}\left(\frac{\sigma_{e0}/100}{M/10^7M_\odot}\right)^{5/14}. \label{eq:nut_sigmaM_scaling}
\end{align}

As demonstrated in Figure~\ref{fig:sum-peaks}, the peaks from each height's turnover frequency add together to form an intermediate regime of the total flux that has a power-law index significantly flatter than $5/2$.
In this example, $M=10^8M_\odot$, $B_0=10^3$, $b=1$, $p=2$, $R_c=R_w=10$, and $H_w/z_0=10^5$.

\begin{figure*}
    \centering
    \includegraphics[width=0.9\textwidth]{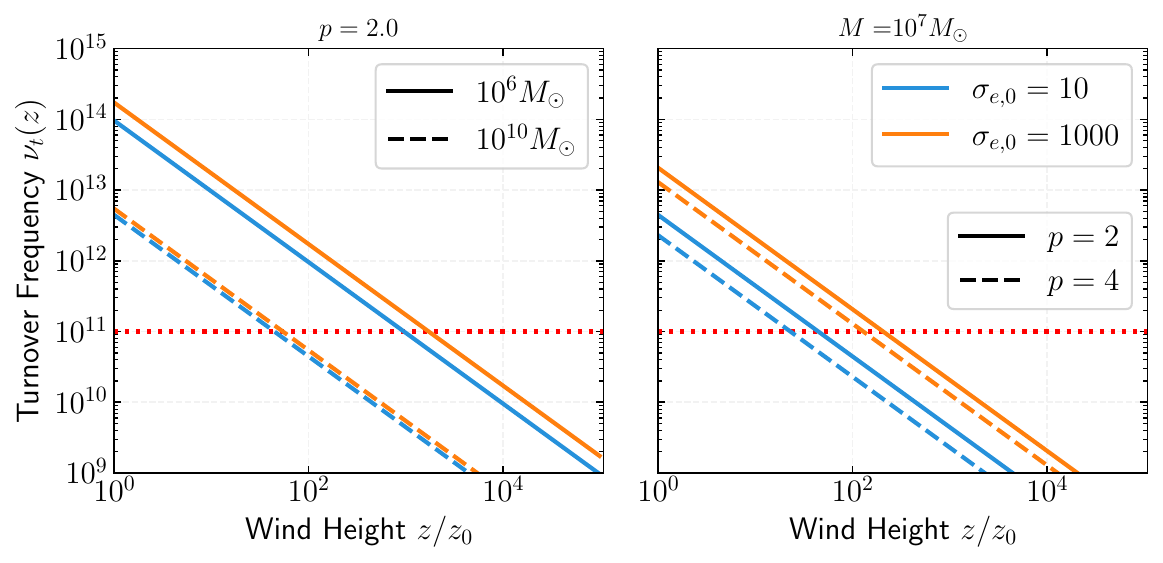}
    \caption{Dependence of the outflow turnover frequency as a function of height in the outflow. The left and right panels show dependence of $\nu_t(z)$ \added{in Hz} on the black hole mass and electron power-law index, respectively. In both panels, colors show different values of the electron magnetization at the outflow base $\sigma_{e0}$. See equation~\eqref{eq:nut_sigmaM_scaling} and~\ref{eq:nut_z_scaling}. Horizontal dotted red lines show where the outflow emits $100~{\rm GHz}$ radiation.}
    \label{fig:nut_properties}
\end{figure*}

The turnover frequency as a function of outflow height is shown in Figure~\ref{fig:nut_properties}.
The turnover frequency can drop by five orders of magnitude from the outflow height to the outflow base, depending on the power-law index, black hole mass, and initial electron magnetization.
The 100 GHz emission tends to come from heights of $10^2-10^3z_0$, i.e. $10^3-10^4r_g$ from the black hole.
The black hole mass sets an absolute scale for the problem, thereby changing the magnitude of the turnover frequency but not its dependence with outflow height (Figure~\ref{fig:nut_properties} left).
Changing the electron power-law index $p$ does not change the slope of the turnover frequency with height (Figure~\ref{fig:nut_properties} right); see equation~\eqref{eq:nut_z_scaling}.

The requirement that 100 GHz emission comes from the expanding outflow sets a lower bound on the outflow extent $H_w/z_0$.
We define the outflow height $z_{100}$ such that $\nu_t(z_{100})\equiv\nu_{100}=100~{\rm GHz}$, i.e. the height at which the turnover frequency is 100 GHz.
For this constant power-law model, the required extent of the outflow is not constraining.
Inverting equation~\eqref{eq:nut_z_scaling} and setting $\nu_t(z_{100})=100~{\rm GHz}$ yields
\begin{align}
    \frac{z_{100}}{z_0}=\left(\frac{100~{\rm GHz}}{\nu_{t0}}\right)^{-1/k_\nu}~&\stackrel{\mathclap{\mbox{p=2}}}{=}~10^3\left(\frac{\sigma_{e0}/100}{M/10^7M_\odot}\right)^{1/3}\label{eq:z100}\\
    &\stackrel{\mathclap{\mbox{p=3}}}{=}~500\left(\frac{\sigma_{e0}/100}{M/10^7M_\odot}\right)^{5/14}
\end{align}
From these equations, we see that the outflow at most needs to extend in height out to $\sim 2\times10^3z_0$ for the upper bound of $p=2$, $\sigma_{e0}=10^3$, and $M=10^6 M_\odot$.
Requiring that $\sigma_e(z_{100})$ be greater than 1 to maintain efficient particle acceleration shows that the magnetization cannot drop off faster than $1/z$.

\subsection{Analytic Flux Scalings}
To obtain analytic scalings of the flux spectral slope and amplitude with parameters such as magnetic field, we assume that only the optically thick part of each outflow height will contribute to the total spectrum.
In this sense, we write $I_\nu(z)=S_\nu(z)\Theta(\nu-\nu_t(z))$ where $\Theta$ is the Heaviside step function and $S_\nu=j_\nu/\alpha_\nu$ is the source function, the optically thick limit of intensity (expression given in Appendix~\ref{app:sourcefunc}).
Under this assumption, the contribution to the flux at a given frequency $\nu_c$ comprises emission from the outflow base up to a height $z_c$, where $\nu_t(z_c)=\nu_c$.
Dropping all constants and substituting equation~\eqref{eq:nut_z_scaling}'s dependency $\frac{z_c}{z_0}= \left(\frac{\nu_c}{\nu_0}\right)^{-1/k_\nu}$, we have the flux $F_{\nu_c}$ at $\nu_c$:
\begin{align}
    F_{\nu_c}&\equiv F_{\nu_c0} \nu_c^{-\alpha_F} \sigma_{e0}^{\beta_F} M^{\gamma_F}\label{eq:fnuscaling}\\
    &\sim \int_{z_0}^{z_c(\nu_c)} S_\nu(z')R_w(z')~{\rm d}z'
    \sim \nu_c^{5/2} B_0^{-1/2}R_{w0}z_0\left(\frac{z_c}{z_0}\right)^{b/2+w+1}\label{eq:fluxscaling}
\end{align}
Because the source function $S_\nu$ is independent of particle density, the exponent $a$ appears only through the turnover frequency scalings rather than the integration over the outflow height. 
We substitute magnetization for magnetic field (equation~\eqref{eq:sigmae}), density for black hole mass (equation~\eqref{eq:n0tau}), and assume that $z_0=R_c=R_{w0}=10r_g\sim M$.
We write this expression in terms of $\sigma_{e0}$ because unlike the magnetic field $B_0$, the magnetization at the outflow base $\sigma_{e0}$ can plausibly be considered constant across different galaxies, as can $\eta_0$ if it, as we assume, depends on $\sigma_{e0}$.
The constant $F_{\nu_c0}$ contains information on the absolute magnitude of $F_\nu$ and depends on parameters at the outflow base and the slopes of the density and magnetization profiles.
The coefficients $\alpha_F$, $\beta_F$, and $\gamma_F$, defined by equation~\eqref{eq:fnuscaling}, are given by:
\begin{align}
    \alpha_F&=-\frac52+\frac{(b/2+w+1)}{k_\nu}\to0\label{eq:alphaF}\\
    \beta_F&=-\frac14+\frac{(b+2w+2)(p+2)}{4k_\nu(p+4)}\to\frac{2p+3}{2(p+4)}\label{eq:betaF}\\
    \gamma_F&=2-\beta_F \to \frac{2p+13}{2(p+4)}\label{eq:gammaF}
\end{align}
where the arrow takes $b=1$ and $a=2$, and $w=1$. 
Notably, for this set of parameters, the spectral slope is flat.
Taking $p=2$ gives $\beta_F=7/12=0.6$, and $\gamma_F=17/12=1.4$.
Setting $D=10~{\rm Mpc}$ and $\eta_0=0.01$, these parameters yield:
\begin{align}
    F_{\nu_c}&=800~{\rm mJy}~\left(\frac{\sigma_{e0}}{100}\right)^{7/12}\left(\frac{M}{10^7M_\odot}\right)^{17/12}.\label{eq:Fnuc}
\end{align}
With $L_{{\rm mm}}=\nu_cF_{\nu_c}\times 4\pi D^2$ with $\nu_c=100~{\rm GHz}$, the ratio between X-ray (equation~\eqref{eq:LX}) and mm luminosity is given by:
\begin{align}
    \frac{L_{\rm mm}}{L_X}&\ge\frac{4\pi D^2\nu_c F_{\nu_c}}{\eta_XB_0^2R_c^2c/2}\sim \nu^{-\alpha_F+1}\left(\frac{\sigma_{e0}}{M}\right)^{\beta_F-1}\\
    &\to3\times10^{-4}\left(\frac{\sigma_{e0}/100}{M/10^7M_\odot}\right)^{-5/12}\left(\frac{\eta_X}{0.1}\right)^{-1}\label{eq:LmmLXratio}
\end{align}
where the last line sets $b=1$, $a=2$, $p=2$.
These analytic scalings will be compared to observational constraints in Sec.~\ref{sec:obs}.
If parameters such as $B$ (and hence $\sigma_{e0}$) and $M$ are independently constrained, then the flux could be used as a way to measure $p$ and thus constrain dissipation profiles in the outflow.

A simple estimate of $L_{\rm mm}/{L_X}$ can also be obtained in a more intuitive way.
Because of the fast cooling of the 100 GHz-emitting electrons (Sec.~\ref{ssec:cooling}), $L_{\rm mm}$ should be approximately equal to the energy contained in the 100 GHz-emitting electrons, or slightly less as the cooling slows down with distance from the corona. 
Setting $L_{\rm mm}=U_c(z) 4\pi R_{w}^2(z) c$ where $U_c$ is the energy density in the $\nu_c=100$ GHz-emitting electrons, we can write $L_{\rm mm}$ in terms of quantities at the wind base because $n(z)R_w^2(z)$ is constant, i.e. $L_{\rm mm}=U_c(z_0)4\pi R_{w0}^2c$.
Combining with the X-ray luminosity, which is determined by the compact corona properties, we have:
\begin{align}
    \frac{L_{\rm mm}}{L_X}&\lesssim\frac{U_c 4\pi R_{w0}^2 c}{\eta_X (B_0^2/8\pi)4\pi R_c^2 c}=\frac{2\eta_0}{\eta_X}\eta_{Ec}\frac{\langle \gamma_{\rm PL}\rangle}{\sigma_e}\\
    &\approx 2\times10^{-4}\left(\frac{\sigma_e}{100}\right)^{-1}\left(\frac{\eta_{Ec}}{3\times10^{-3}}\right),\label{eq:Uc}
\end{align}
where we have defined $\langle \gamma_{\rm PL}\rangle\equiv U_{\rm PL}/(n_{\rm PL}m_ec^2)$, with $U_{\rm PL}=\int_{\gamma_1}^{\gamma_2}\gamma m_ec^2 N_{\rm PL}d\gamma$ using (equation~\eqref{eq:nPL}) is the energy density in the entire power-law.
For typical values of $p=3$, $\langle\gamma_{\rm PL}\rangle\approx5$.
For our chosen parameters, most of the electron energy is contained in the power-law component: $U_{\rm PL}/(U_{\rm PL} + U_{\rm MJ})=(\langle\gamma_{\rm PL}\rangle/\theta_e) \eta_0/(1-\eta_0)\approx0.74$, where $U_{\rm MJ}=n_{\rm MJ}k_BT_e$ is the energy density of the thermal electrons.
Therefore, we define the fraction of energy contained in the 100 GHz-emitting electrons compared to the total particle energy as $\eta_{Ec}=U_c/U_{\rm tot}\approx U_c/U_{\rm PL}$.
Explicitly, $\eta_{Ec}=(\int_{\gamma_c}^{\gamma_c+\Delta\gamma}\gamma N_{\rm PL}d\gamma)/(\int_{\gamma_2}^{\gamma_1}\gamma N_{\rm PL}d\gamma)$, where $\gamma_c\approx100$ according to equation~\eqref{eq:nus}.
For the nonthermal number fraction $\eta_0$ it is reasonable to assume that $\eta_0\approx\eta_X$, and so equation~\eqref{eq:Uc} can be understood as a relation between $L_{\rm mm}/{L_X}$ and the kinetic energy fraction of the 100 GHz-emitting nonthermal electrons.


\section{Example Outflow Properties: GRMHD Simulations}\label{sec:grmhd}
To provide one example of how an outflow connected to the X-ray corona might appear, we consider a general relativistic magneto-hydrodynamic simulation of an accretion disc.
This example is largely for illustrative purposes, to demonstrate the presence of such an outflow and how it relates to other components of the accretion disc-black hole system.
Future work will pinpoint the exact location of the outflow relative to the corona and disc. 
To focus on luminous AGN, we examine a simulation of a thin accretion disc, similar to the canonical~\citet{shakurasunyaev} model.
Because the goal is to find a set-up analogous to radio-quiet AGN without strong jets, we select a simulation with a magnetic field configuration that does not lead to a relativistic jet in the polar regions, in contrast to magnetically-arrested discs. 
In particular, we analyze the RADTOR model from~\citet{liska2022}, a high resolution simulation that was initialized with a purely toroidal magnetic field with nearly uniform ratio of plasma gas pressure to magnetic pressure $\beta\sim7$.
This simulation, first presented in~\citet{liska2022}, was performed using the H-AMR code~\citep{hamr}. 
\added{Through a combination of static mesh refinement and adaptive mesh refinement, the effective resolution at the midplane was 4080 × 1728 × 1152 in the $r$, $\theta$, and $\phi$ dimensions, respectively.
The simulation domain included $\pi$ radians in $\theta$, $2\pi$ radians in $\phi$, and extended from the event horizon to $10^4r_g$.}
The simulation utilizes two-temperature thermodynamics and M1 radiation transport to evolve the gas around a rapidly-spinning black hole ($a=0.9375$). 
The resulting disc has a scale height to radius ratio $H/R=0.03$; see~\citet{liska2022} for full simulation details.
Quantities are in code units unless otherwise specified.
  
\begin{figure*}
    \centering
    \includegraphics[width=0.95\linewidth]{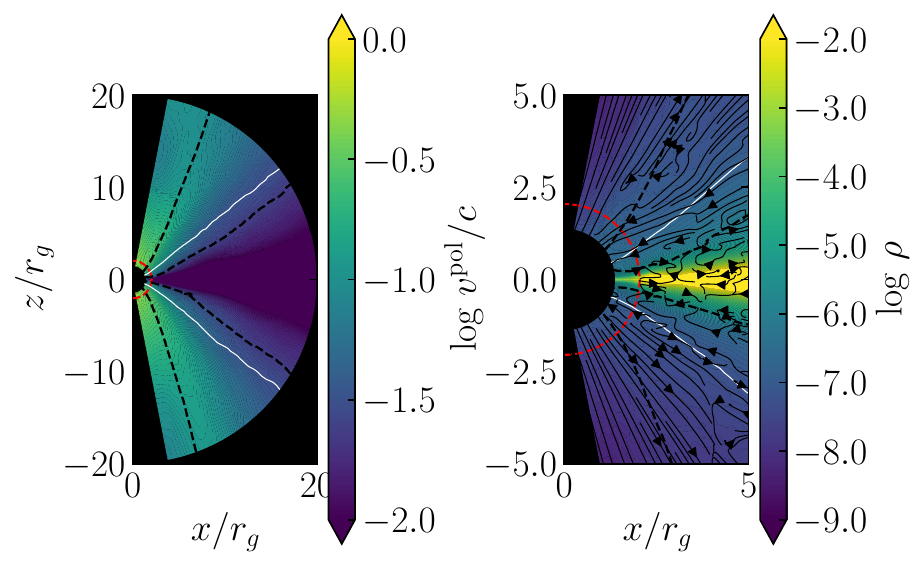}
    \caption{Vertical slice of the radiative GRMHD simulation of a thin disc showing the poloidal 3-velocity (equation~\eqref{eq:vpol}). 
    The ``outflow'' refers to the region inside the dashed black contours marking $\sigma_h\in[10^{-2},2]$.
    The right panel shows a zoom-in of the same slice colored according to mass density, with magnetic field lines in black.
    The white line shows where the Bernouilli constant is zero, such that material above this line is unbound. The red dashed line shows the value of the equatorial innermost stable circular orbit. 
    These slices have been azimuthally-averaged and time-averaged over $17.5\times10^4 <tc/r_g<17.9\times10^4$.}
    \label{fig:slice}
\end{figure*}

The inner disc region features the outflow, as shown by Figure~\ref{fig:slice}. 
This outflow has a relatively fast poloidal velocity of $0.2c$ close to the black hole that drops to about $0.06c$ at $50r_g$. 
Here we have defined the poloidal velocity $v^{\rm pol}$ as:
\begin{equation}
    v^{\rm pol}\equiv\sqrt{(v^r)^2+(v^\theta)^2}    \label{eq:vpol}
\end{equation}
where $v^r=u^r/u^t$ and $v^\theta=u^\theta/u^t$.
The right panel of Figure~\ref{fig:slice} shows a zoom-in with magnetic field lines in black, showing how the outflow connects to the inner region of the black hole.
\added{We note that this figure shows only a subset of the full simulation domain.}

The outflow region is selected by cuts in the plasma magnetization $\sigma_h\approx\sigma_i$, as shown by the black contours in Figure~\ref{fig:slice}.
For this simulation, the outflow is between $0.01<\sigma_h<2$; note that the electron magnetization values $\sigma_e$ will be a factor of $\approx m_p/m_e$ higher than $\sigma_h$.
We select only regions between $20^\circ<\theta<80^\circ$ to avoid the disc body and atmosphere.
The properties of the outflow do not change significantly if the upper/lower limits of the cuts are varied slightly, for example between $0.1<\sigma_h<1$ or including slightly lower/higher $\theta$ values. 
We also show where the gas becomes unbound through a contour of the Bernoulli constant $-u_th/\rho-1$, where $h=\rho+\gamma P/(\gamma-1) + b^\mu b_\mu$ is the gas enthalpy, $P$ is the gas pressure, and $b^\mu b_\mu$ is the magnetic energy density.

\begin{figure}
    \centering
    \includegraphics[width=\linewidth]{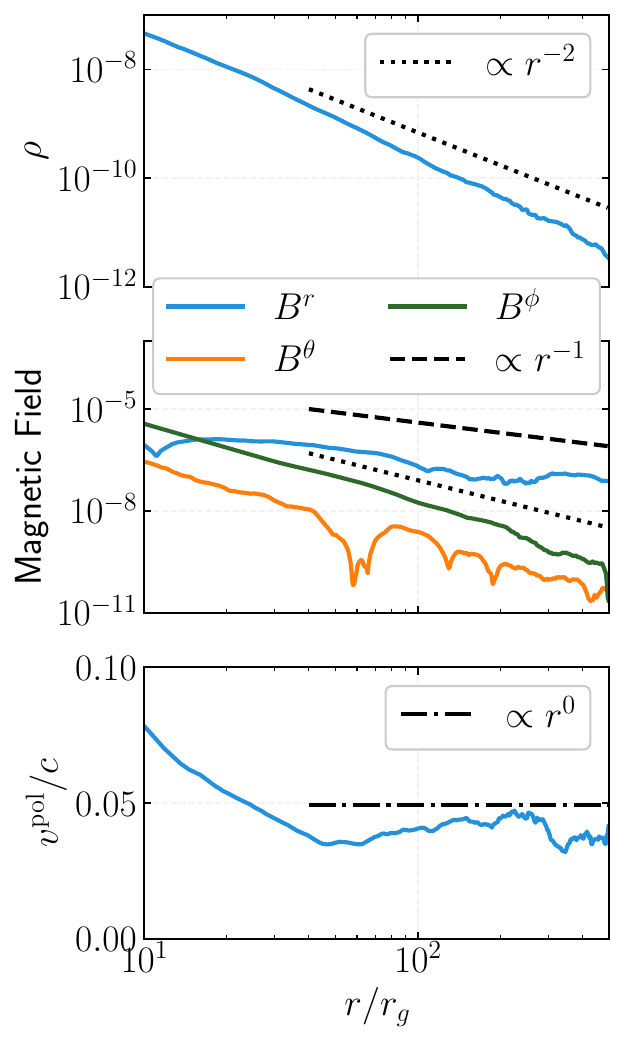}
    \caption{Profiles of mass density $\rho$, magnetic field components $B^r$, $B^\theta$, and $B^\phi$, and poloidal three-velocity $v^{\rm pol}$ in the simulated outflow, averaged over azimuth and outflow latitude.
    Power laws are shown as black dotted lines ($\propto r^{-2}$), dashed lines ($\propto r^{-1}$), and dash-dot lines ($\propto$ const.).
    These profiles have been time-averaged over $17.5\times10^4 <tc/r_g<17.9\times10^4$.}
    \label{fig:outflowprofiles}
\end{figure}

The one-dimensional properties of the outflow are extracted by averaging over azimuth and latitude once the 3D outflow region has been selected.
The resulting radial profiles in density and magnetic field are shown in Figure~\ref{fig:outflowprofiles}.
After non-outflow-related behavior close to the black hole, the outflow reaches steady state properties for $r\gtrsim50r_g$.
Power-law fits to these quantities yield approximately $\rho\sim r^{-2}$ and $B\sim r^{-1}$, motivating the use of $a=2$ and $b=1$ in the outflow model.

Although this simulation does not have a strong jet due to the lack of strong poloidal flux, the outflow is still present in simulations of magnetically-arrested tori and discs, e.g. RADPOL in~\citet{liska2022} (not shown). 
We note that failed winds are present just below the outflow itself and above the disk body (yellow), as visible in the curved magnetic field lines.
These failed winds are approximately the same density as what we envision the corona to be, whereas the outflow from the corona has a density approximately ten times lower.

\section{Observational Implications and Constraints}\label{sec:obs}
\begin{figure*}
    \centering
    \includegraphics[width=0.95\linewidth]{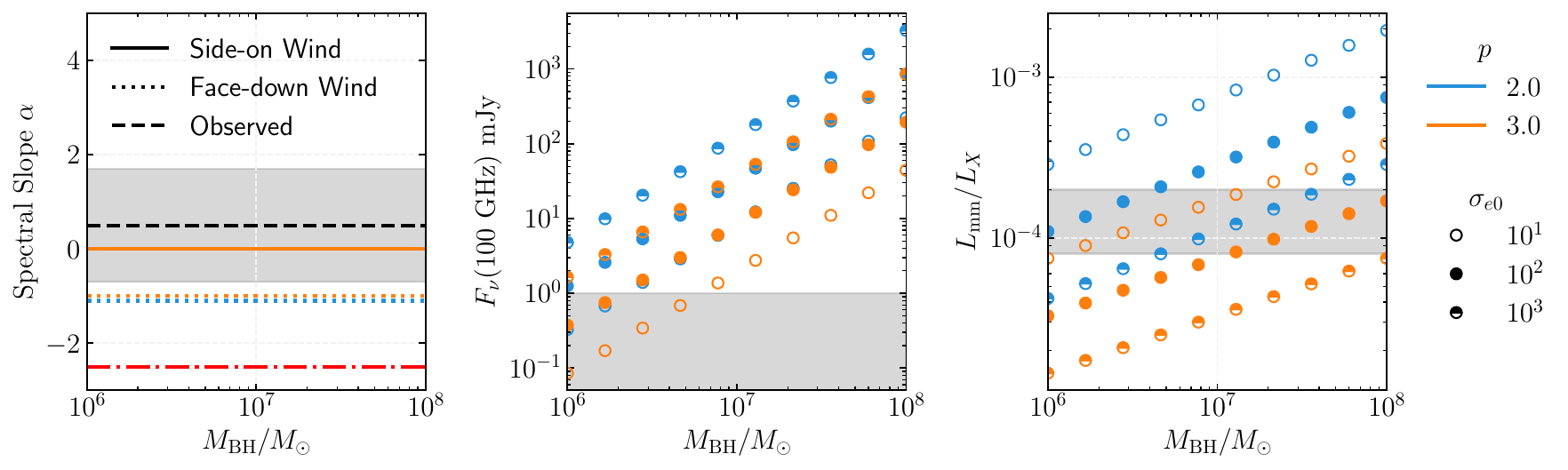}
    \caption{Observational properties of the outflow as a function of black hole mass. 
    Left: the spectral slope $\alpha$ such that $F_\nu\sim\nu^{-\alpha}$ at 100 GHz for $a=2$, $b=1$ as predicted for a side-on outflow (equation~\eqref{eq:alphaF}) and face-down outflow (Appendix~\ref{app:facedown}). The gray shaded area indicates the spread in observed spectral slopes, with the black dashed line showing the average observed value. The red dash-dot line shows the value $\alpha=-5/2$ predicted for optically-thick emission. 
    Middle: the flux at 100 GHz predicted for the side-on outflow model (equation~\eqref{eq:Fnuc}).
    The gray region shows approximately the limit of observations.
    Right: the predicted ratio between mm and X-ray luminosity for the side-on outflow model (equation~\eqref{eq:LmmLXratio}). The gray region shows the observed spread $\log(L_{\rm mm}/L_{\rm X})=7\times10^{-4}\pm3\times10^{-4}$~\citep{kawamuro2022}.
    }
    \label{fig:obs-over-M}
\end{figure*}
In this section, we explore the model parameter space that is consistent with observational constraints for the compact millimeter emission in RQAGN.
These constraints are:
\begin{itemize}
    \item $F_\nu\sim\nu^{-\alpha_{\rm mm}}$ with $\alpha_{\rm mm}=0.5\pm1.2$~\citep{kawamuro2022}. 
    \item $L_{\rm mm}/L_X$ is roughly constant across three orders of magnitude in black hole mass~\citep{ricci2023}.
    \item $L_{\rm mm}/L_X\sim 10^{-4},~10^{-5}$~\citep{behar2015, kawamuro2022, ricci2023}.
    \item $L_{\rm mm}/L_X$ is roughly inclination-angle independent~\citep{kawamuro2022}.
\end{itemize}
The spectral slope $\alpha_{\rm mm}$ constrains the power-law indices of the magnetic field and density, whereas the magnitude of $L_{\rm mm}/L_{\rm X}$ constrains the conditions at the outflow base through $F_{\nu0}$ (equation~\eqref{eq:Fnuc}) and $L_X$ (equation~\eqref{eq:LX}). 
The spectral slope's lack of dependence on inclination angle suggests that the emission is unlikely to come from a highly-beamed source such as a jet.

Figure~\ref{fig:obs-over-M} shows the model constraints as a function of black hole mass.
The first observational constraint directly applies to $\alpha_F$, as shown in the left panel. 
Because the spectral slope is entirely determined by the power-law indices of the magnetic field and density, $\alpha_F$ is independent of the black hole mass. 
The model's predicted slope is between 0 and around -1, depending on the viewing angle of the outflow and the outflow's geometry.
In this plot, the side-on outflow has a conical profile ($w=1$) whereas the face-down outflow has a parabolic profile ($w=0.5$).
The conical outflow's spectral slope does not change with viewing angle and is always 0.
The parabolic outflow's spectral slope is weakly dependent on viewing angle, with a spectral slope of $\alpha=-1$ for the face-down limit (Appendix~\ref{app:facedown}) and $\alpha=-(p+9)/2(p+3)$ for the side-on viewing limit (equation~\eqref{eq:alphaF}).
Although the face-down parabolic outflow's spectral slope is slightly below the observed range $\alpha\approx0.5\pm1.2$~\citep{kawamuro2022}, it still maintains a value notably flatter than the optically-thick limit (shown as a red dash-dot line).
The conical outflow always lies within the observed range and its lack of dependence on viewing angle is consistent with~\citet{kawamuro2022}.

As per equation~\eqref{eq:Fnuc}, the flux at 100 GHz scales with the black hole mass approximately with $M^{1.4}$ (middle panel). 
The flux depends weakly on the magnetization at the outflow base $\sigma_{e0}$, increasing by approximately an order of magnitude with two orders of magnitude increase in $\sigma_{e0}$.
Increasing the electron power-law index $p$ from 2 to 3 leads to about a factor of 2 increase in the 100 GHz flux. 
The observed 100 GHz flux is typically around 1 mJy~\citep{kawamuro2022}, indicating that something likely decreases the observed flux from its predicted intrinsic value. 

The mass-independence of $L_{\rm mm}/L_X$ in principle constrains $\gamma_F$ (equation~\eqref{eq:gammaF}).
However, as equation~\eqref{eq:LmmLXratio} shows, if $\sigma_{e0}$ depends on black hole mass, then the ratio changes.
For example, a fixed ratio $\sigma_{e0}/M$ would yield $L_{\rm mm}/L_X$ independent of mass; interestingly, this ratio suggests a constant magnetic field $B$ as a function of black hole mass. 
To demonstrate the range of possible values, Figure~\ref{fig:obs-over-M}'s right panel shows both the spread in $L_{\rm mm}/L_X$ and the magnitude for a range of $\sigma_{e0}$ and electron power-law index $p$.
The ratio between millimeter and X-ray flux clusters around the observed value of $10^{-4}$, $10^{-5}$, with spread according to $p$ and $\sigma_{e0}$, although changing $p$ from reasonable values of 2 to 3 changes the ratio by about a factor of 2. 
The predicted ratio approximately matches the observed value.
The magnitude and spread replicate e.g.~\citet{ricci2023}'s Figure 2 relatively well, further demonstrating the applicability of the extended outflow model.

\section{Discussion}
\subsection{Dissipation Mechanisms Along the Outflow}
One hope for detecting mm emission from the coronal outflow is that it can be used to elucidate whether magnetic reconnection, turbulence, or shocks dominates the conversion of magnetic energy into heat and nonthermal particle energy. 
The power-law index $p$ could in principle probe these mechanisms since it depends on acceleration mechanism and plasma properties~\citep{werner2018, zhdankin2020}. 
However, the conical outflow model appears too robust with $p$ to discriminate between $p=2$ and $p=3$, the commonly accepted range of power-law indices.
The spectral slope is completely independent of $p$ (equation~\eqref{eq:alphaF}), and the mm flux (and therefore luminosity) changes by about an order of magnitude with $p$ going from 2 to 3 (Figure~\ref{fig:obs-over-M}), which is not within current observations' ability to resolve. 
Accordingly, alternative probes of dissipation mechanisms in the corona are needed, potentially through studying the characteristic variability of a plasma dominated by reconnection vs. turbulent dissipation.

Although strong magnetizations could in principle lead to proton acceleration as well~\citep{mbarek2024, fiorillo2024}, we neglect these protons because they will not contribute to the synchrotron emission.
Any protons accelerated in the outflow will be unlikely to contribute to neutrino flux since the X-ray energy density drops as $1/r^2$ away from the corona.

\subsubsection{Sub- and Super-Alfv\'enic Dissipation} \label{sssec:superalfvenic}
\added{One certainty resulting from the extended outflow model is that, \added{due to the short cooling times discussed in Sec.~\ref{ssec:cooling},} the electrons must be continually accelerated within the body of the outflow rather than advected from the $10r_g$ region.
Although continual dissipation has been assumed in jet models for decades~\citep[e.g.]{bk}, the exact mechanisms of this dissipation remain unclear.
Toy models suggest that dissipation can occur at large distances from the black hole~\citep{mehlhaff2025}.
In principle, dissipation due to the cascade of Alfv\'en waves should be sufficient to produce the proposed electron distribution below the Alfv\'en point.
However, the exact mechanism for producing counter-propagating Alfv\'en waves, the impact of density gradients on the development of the turbulent cascade, and the energy balance between the Alfv\'enic and fast mode cascade remains unclear.
Future work will probe the nature of dissipation in the outflow.}

The outflow is expected to be roughly sub-Alfv\'enic at the height $z_c$ that dominates 100 GHz emission to effectively dissipate energy into nonthermal particles.
The relevant equation for relativistic Alfv\'en speed of the outflow is given by
\begin{equation}
  v_A(z)=c\sqrt{\frac{\sigma_p(z)}{\sigma_p(z)+1}}\to v_{A0} \\
\end{equation}
where
\begin{align}
v_{A0}&\approx 0.2\left(\frac{\sigma_{e0}/100}{\sigma_{e0}/100 + m_p/m_e}\right)^{1/2}c. \label{eq:vA0}
\end{align}
The outflow will become sub-Alfv\'enic at a height $z_A$, where $v_z(z_A)=v_A$.
For an initial \added{outflow speed of $v_{z0}=0.002c$ and assuming a radial dependence given by equation~\eqref{eq:vz}, the Alfv\'en point occurs around $100z_0$, which is an order of magnitude or more below the height $z_{100}$ where the 100 GHz emission occurs (equation~\ref{eq:z100}).}
However, we note that if mass flux is not conserved, $v_z$ could drop with height as well.
We also note that other components of the magnetic field could increase the Alfv\'en speed. 
These two effects could push the Alfv\'en point to larger radii. 

\added{Although the bulk kinetic energy dominates the magnetic energy above the Alfv\'en point, it is currently unclear whether magnetic dissipation processes will turn off or become subdominant to other mechanisms. 
In the solar wind, dissipation continues efficiently beyond the Alfv\'en point~\citep{bandyopadhyay2023}, perhaps due to ion-cyclotron processes~\citep{bowen2022} or large-scale shear~\citep{ruffolo2020}.
A similar effect could occur in this outflow, potentially decreasing the efficiency of converting magnetic energy into particle energy.
We leave detailed modelling of the dissipation mechanisms along the outflow to future work.}

\subsection{Connecting the Accretion Disc and Corona}
We have neglected any direct coupling of the accretion disc and the corona, \added{and in particular have not assumed any geometry or magnetic field configuration in the corona beyond its size and magnetic field strength}.
As such, we do not directly address the heating problem of the corona, instead assuming that outflows such as the ones seen in GRMHD simulations exist and have properties similar to those outlined in Sec.~\ref{sec:grmhd}.
We leave constraining the exact outflow-launching mechanism to future work.

These outflows are present in GRMHD simulations across accretion disc types -- thin vs. thick, magnetically-arrested (MAD) vs. weakly magnetized -- though disentangling them from jets can be quite difficult~\citep{scepi2024, manikantan2024, dhang2025}. 
In addition to the simulation discussed in Sec.~\ref{sec:grmhd} from~\citet{liska2022}, thin MAD discs also show Blandford-Payne type winds~\citep{avara2016, scepi2024}. 
Indeed, thin MADs could have more powerful winds per unit jet power than thick MADs~\citep{avara2016}.
Fully characterizing the differences between winds from truncated MADs and pure Shakura-Sunyaev-type thin accretion discs is beyond the scope of this work. 
Although GRMHD simulations could in theory be used to constrain the outflow geometry and thus the $w$ parameter (equation~\eqref{eq:Rwz}), in practice reaching equilibrium out to $10^3r_g$ scales or larger is computationally challenging.

The extended outflow is fundamentally separate from a relativistic jet driven by extracting a black hole's rotational energy~\citep{bz}. 
The extended outflow in principle still exists even in the presence of a jet, for example in the RADPOL simulation of~\citet{liska2022} (not shown).
Although optically-thin synchrotron emission likely dominates the flat spectrum of the outflow, interactions of the jet and outflow could be a source of heating and nonthermal particle acceleration through e.g. Kelvin-Helmholtz instability~\citep{sironi2021, tsung2025} or dissipation of the imbalanced Alfv\'enic turbulence along the jet-outflow boundary~\citep{hankla2022a}. 
\added{Previous models of MHD winds have found that the ionization structure agrees well with X-ray observations~\citep{fukumura2010}.}


\subsection{Comparison to Observations}
The analytic model presented in this work calculates only intrinsic values for e.g. X-ray and mm luminosities.
We do not include any obscuration by line-of-sight effects or from sources on scales larger than $\approx 10^4r_g$, such as a weak jet, star formation, or dust. 
Due to the difficulty of observing on sub-millipc scales, observations of the mm emission likely contain light from sources other than the compact corona. 
Including these effects could change the spectral slope, mm flux, and luminosity ratio $L_{\rm mm}/L_X$.
State-of-the-art observations disagree on the dominant source of emission at 100 GHz, with some suggesting free-free emission~\citep{mutie2025}.
Even models that find an excess in the mm above other, larger-scale sources can have a significant fraction of flux from optically-thin jet synchrotron emission extrapolated from lower frequencies~\citep{shablovinskaya2024}. 
The most direct measurement of the mm emitting region's size estimated a spectral slope of $\alpha=0.26\pm0.4$ and has $L_{\rm mm}/L_X=10^{-4.3}$~\citep{rybak2025}, in agreement with our model predictions.
Although our extended outflow model predicts a flat spectrum up to $\sim10^{14}$ Hz, dust and other effects likely dominate at those higher frequencies. 

\subsubsection{Observational Evidence for an Extended mm-emitting Region}
Testing the claim of synchrotron emission from the corona or the outflow is difficult observationally because of the large angular resolution required. 
Without micro-arcsecond angular resolution, larger-scale sources of light will almost certainly contaminate coronal emission. 
Resolving the mm region would require a spatial resolution of $50-10^4r_g$, depending on the inclination angle: for a quasar such as 3C 273 located 760 Mpc away with a black hole mass of $3\times10^8M_\odot$~\citep{gravity2018}, this spatial resolution corresponds to an angular resolution of $0.4$ - 80 milli-arcseconds.  

Because spatially resolving the corona remains challenging for most black holes, most observational attempts to measure the size of the mm-emitting region rely on estimating light-crossing times from the variability of the mm emission on day-long timescales.
These methods yield size estimates of $\lesssim300r_g$ for the mm-emitting region~\citep{michiyama2024, shablovinskaya2024}.
We note that these methods measure the size of the emitting structures rather than their location relative to the central black hole; in particular, their constraints of $\approx100 r_g$ sizes could come from a $100r_g$ size blob far away from the central black hole. 
We also note that some works have suggested an extended X-ray corona with sizes upwards of $250r_g$~\citep{palacio2025}.
In particular, simultaneous X-ray and mm emission studies have suggested that mm comes from $\lesssim1500r_g$, proposing that the two regions are connected but not spatially co-located~\citep{petrucci2023}.

A more robust constraint on the size of the mm-emitting region comes from the recent observation of a microlensed quasar~\citep{rybak2025}.
Their constraint of $\lesssim100r_g$ comes from measuring flux ratios of the four gravitationally-lensed images. 
The compactness could originate from projection effects resulting from the non-spherical geometry of our extended outflow geometry, as demonstrated in Appendix~\ref{app:facedown}'s Figure~\ref{fig:facedown_diagram}.
In particular, if the outflow is viewed face-down, the 100 GHz emission appears to come from $\approx10R_{w0}\to100r_g$ away from the central black hole (Figure~\ref{fig:nut_facedown}), in agreement with the microlensing result. 

We also note that the extended outflow model gives a flat spectrum over a wide range of frequencies, whereas the uniform spherical model will only have a flat spectrum at a narrow range of frequencies perfectly around the turnover frequency. 
In this sense, there is a fine-tuning argument against the uniform spherical model; why does the observed frequency just happen to be very nearly the turnover frequency?
An extended mm-emitting region is robust against this fine-tuning argument.

\subsubsection{Comparison to Previous Models}
\begin{figure}
    \centering
    \includegraphics[width=0.95\linewidth]{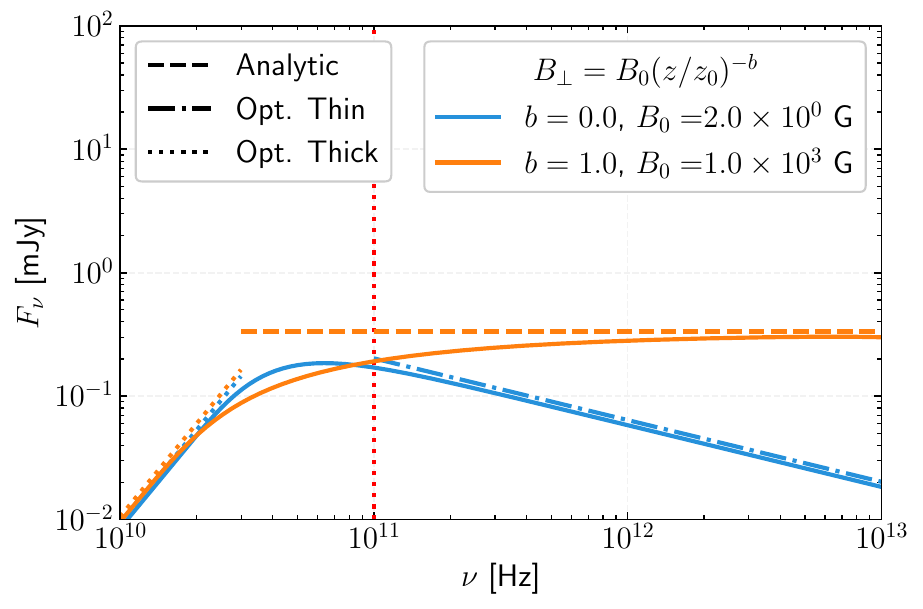}
    \caption{Assumptions on the magnetic field profile strongly affect predictions for mm flux. 
    For a given flux at 100 GHz of 0.1 mJy, the commonly-assumed model with a homogeneous magnetic field profile (blue) yields horizon-scale fields three orders of magnitude lower than the decaying magnetic field proposed in this work (orange).
    The only difference in these two models is the magnetic field strength $B_0$ at $10r_g$ and the slope $b$ of the magnetic field's decay.}
    \label{fig:bslope_demo}
\end{figure}
As discussed in Sec.~\ref{ssec:stronglymag}, models relying on the assumption of a spherical, homogeneous mm-emitting region often calculate magnetic field strengths on the order of 1 - 10 G by inverting a variant of equation~\eqref{eq:nutsphere} that assumes the peak mm flux is the optically-thick synchrotron turnover frequency~\citep{inoue2018, shablovinskaya2024}.
As Figure~\ref{fig:bslope_demo} demonstrates, the difference in fitted magnetic field is due to assumptions on the magnetic field profile.
This figure takes a conical outflow model with all parameters the same (e.g. $n(z)\sim z^{-2}$, $p=2$) except for the magnetic field power-law index and the initial value of magnetic field. 
The results yield equivalent flux at 100 GHz, but the models fit magnetic fields of 2 G and $10^3$ G, the former value being too low for the corona to be magnetically-powered whereas the latter is consistent with equation~\eqref{eq:LX}. 
The extended outflow model therefore reproduces the qualitative observational characteristics while also invoking a physically-motivated electron distribution. 

\added{The closest idea to this work is probably the non-uniform model of~\citet{raginski2016}, which considered a sandwich-type corona sitting in a slab geometry above the accretion disk as well as a spherical geometry. 
In both cases, the assumption of equipartition between nonthermal electrons, thermal electrons, and the magnetic field leads to lower magnetic field values in the X-ray emitting region than those argued for in this work; the magnetic field values in the mm-emitting region are comparable.
This equipartition assumption also means that the mm-emitting and X-ray-scattering electrons are spatially co-located, in contrast to this work.
Notably, this work and~\citet{raginski2016} obtain a flat spectral slope in the mm emission via zones of gas with different densities and magnetic fields stacking together as in Fig.~\ref{fig:sum-peaks}.
Both models include a prediction for emission at higher frequencies ($\gtrsim10^{12}$ Hz).}

\subsubsection{Predictions for Simultaneous X-ray and mm Variability}
Recent observational studies have attempted to correlate mm and X-ray variability, since presumably such a correlation would indicate the co-location of the mm-emitting electrons and X-ray-upscattering electrons.
Such correlations have so far been relatively inconclusive, though suggestive of an extended corona~\citep{petrucci2023, shablovinskaya2024}. 
In our extended outflow model, correlations between X-ray and mm are difficult to predict because the mm emission comes from a wide range of length scales.
In principle, variability could come from local changes in e.g. magnetization, or from a perturbation that is advected along the outflow, leading to correlated variability.

We first estimate the impact of changes in magnetization on the mm variability. 
According to equation~\eqref{eq:fnuscaling}, the flux will change from $F_1$ to $F_2=F_1/3$ if the respective magnetizations $\sigma_{e1}$ and $\sigma_{e2}$ change by $F_2/F_1=1/3=(\sigma_2/\sigma_1)^{\beta_F}$, yielding $\sigma_{e1}/\sigma_{e2}=6$ \added{for the canonical value of $\beta_F=0.6$}. 
If we assume this decrease in magnetization is purely due to a decrease in magnetic field, then the synchrotron cooling time increases by the same amount, a factor of 6. 
That means the cooling time would change from $t_{\rm sync,~1}=10^5~{\rm s}(B/4~{\rm G})^{-2}(\gamma/100)^{-1}\approx1~{\rm days}$ to $t_{\rm sync,~2}=6\times10^5~{\rm s}\approx7~{\rm days}$, with $B=4~{\rm G}$ and $\gamma=100$ taken from equation~\eqref{eq:nus}. 
These estimates are quite close to findings that the mm flux varies by a factor of 3 in $\approx 4$ days~\citep{shablovinskaya2024}.

As a crude estimate of the lag between mm and X-ray fluctuations, consider a perturbation in magnetic field strength that originates from close to the black hole, thereby influencing the X-ray emission.
If this perturbation then propagates outward along the outflow at the outflow speed of $\approx0.1c$, it would emit 100 GHz emission from around $10^4r_g$ (Figure~
\ref{fig:nut_properties}).
The lag time between X-ray and mm emission would then correspond to the propagation time $10^4r_g/0.1c$, or
\begin{equation}
    t_{\rm lag}\approx 10^5r_g/c=400~{\rm days}~\left(\frac{M}{10^7M_\odot}\right)
\end{equation}
This estimate depends on the outflow speed and parameters such as the initial magnetization in addition to black hole mass.
It could also be greatly reduced by a variety of factors depending on the nature of the fluctuation.
This large time delay between X-ray and mm could make observationally testing correlations quite difficult.

We also note that a structure that begins as a fluctuation of a given size in the X-ray corona could expand as it propagates along the outflow.
For example, variability from a fluctuation that comes from a light-crossing time of the corona would have a timescale of $10r_g/c$.
By the time this fluctuation propagates to $10^4r_g$, where it would emit mm wavelengths, the radius of the outflow has expanded to $\approx10^4r_g$ (equation~\eqref{eq:Rwz}), thereby increasing the light-crossing time by three orders of magnitude.
This expansion would mean that a day-long variation in the X-rays would correspond to a multi-year variation in the mm, potentially explaining the lack of observed correlation thus far~\citep{shablovinskaya2024}. 
Based on our present understanding, the mm variability is likely related to fluctuations in the cooling time (via the magnetization) rather than propagation effects across an extended outflow connected to the corona.

\subsubsection{Connection to X-ray Binaries}
A model similar to the one presented in this work has been shown to reproduce the scaling between radio and X-ray emission in XRB systems~\citep{kylafis2023, kylafis2024}.
Like ours, this model emphasizes the need for an outflowing corona due to the positive Bernouilli constant predicted by~\citet{adios}.
In contrast to our model, however, the same electrons are Compton upscattering soft photons and radio emission by synchrotron.
This co-location is enabled by assuming that the wind extends over a larger area of the accretion disc, i.e. setting $R_{w0}=100r_g$ or so. 


\section{Conclusions}
In this work, we have argued that both the X-ray emitting corona and the mm-emitting region are magnetically powered.
Following from this physically-motivated assumption, we investigated the consequences of the strong magnetic fields on the geometry of the multiwavelength-emitting region and the electron distribution within that region.
Because a magnetically dominated corona's strong magnetic fields obliterate mm emission from within $10r_g$ of the central black hole due to synchrotron self-absorption, we propose that the mm emission comes from further away, on the order of $10^4r_g$.
Drawing inspiration from GRMHD simulations, we construct a model for an inhomogeneous, extended mm-emitting region that connects to the compact X-ray corona via magnetic fields.
Our proposed multizone, outflowing mm-emitting region has a flat spectrum and millimeter to X-ray luminosity ratio that are relatively independent of viewing angle, in agreement with observations. 
Although the quantities investigated here likely do not enable probing the underlying electron distribution function and thereby the dissipation processes in the corona, our model highlights the role of strong magnetic fields in regions close to black holes. 

Predicting the mm polarization for the extended outflow model is beyond the scope of this work, as it likely requires polarized raytracing of a GRMHD simulation.
\added{Several factors could influence the overall polarization of the mm emission.}
\added{Although} in principle, optically-thin \added{synchrotron} emission in an ordered magnetic field is strongly polarized with linear polarization degrees upwards of 70\%~\citep{rybickilightman}, \added{in practice}, any tangling in the magnetic field will greatly reduce the polarization degree.
In addition, the polarization degree of self-absorbed synchrotron emission is unclear, though presumably is quite weak.
\added{So far, the lack of mm polarization detected from RQAGN is consistent with these expectations~\citep{shablovinskaia2025}.}
\added{In this model,} the total polarization at a given frequency will be a combination of the optically-thin and optically-thick emission from different outflow heights.
Although the optically-thin emission constitutes a small percentage of the total flux, if it is strongly polarized, then it could inflate an otherwise \added{zero} overall polarization degree \added{to modest values consistent with established upper limits of $\lesssim1\%$~\cite{shablovinskaia2025}}.
\added{Beyond mm polarization, the outflow could potentially affect polarization in the X-rays, as observed by the Imaging X-ray Polarimetry Explorer (IXPE;~\citep{ixpe}) in several radio-quiet AGN~\citep[e.g.]{ingram2023, marin2024, gianolli2024}.}
\added{Although} bulk Comptonization off an outflow could boost an otherwise low polarization degree in the X-rays~\citep{poutanen2023, dexter2024}, \added{the required velocities ($\gtrsim0.4c$) are marginally higher than those envisioned in the presented outflow from the corona}.

\begin{acknowledgments}
Support for this work was provided by NASA through the NASA Hubble Fellowship grant \#HF2-51555 awarded by the Space Telescope Science Institute, which is operated by the Association of Universities for Research in Astronomy, Inc., for NASA, under contract NAS5-26555.
This work was supported by NASA grant 80NSSC22K1054 (AP), Simons Foundation (00001470, AP and GM), and facilitated by Multimessenger Plasma Physics Center (MPPC, AMH and AP), NSF grant No. PHY-2206607. 
A.P. additionally acknowledges support by an Alfred P. Sloan Fellowship, and a Packard Foundation Fellowship in Science and Engineering.
GM is supported by a Canadian Institute of Theoretical Astrophysics (CITA) postdoctoral fellowship and acknowledges support from the Simons Collaboration on Extreme Electrodynamics of Compact Sources (SCEECS). 
An award of computer time was provided by the Innovative and Novel Computational Impact on Theory and Experiment (INCITE) and ASCR Leadership Computing Challenge (ALCC) programs under award AST178. This research used resources of the Oak Ridge Leadership Computing Facility, which is a DOE Office of Science User Facility supported under Contract DE-AC05-00OR22725. ML was supported by the John Harvard, ITC and NASA Hubble Fellowship Program fellowships, and NASA ATP award 80NSSC22K0817.
DG is supported by the Research Foundation--Flanders (FWO) Senior Postdoctoral Fellowship 12B1424N.
The authors thank useful discussions with S. Solanki, M. Avara, K. Long, C. Ricci, E. Behar, J. Mehlhaff, and S. Laha. 
\end{acknowledgments}

\begin{contribution}
AMH shaped the research concept, developed the model and analysis, and wrote and submitted the manuscript.
AP shaped the research concept, helped develop the model, obtained funding, and guided the project.
RM contributed to the development of the project idea and edited the manuscript.
RFM provided observational motivation and guidance to the project.
GM provided the GRMHD data and analysis tools.
DG contributed early ideas to the project and edited the manuscript.
ML wrote the \texttt{HAMR} code and ran the GRMHD simulation.
\end{contribution}

\software{HAMR \citep{hamr}}


\appendix
\section{Radiative Transfer Details}
\subsection{Characteristics of Synchrotron Self-Absorption and Emission} \label{app:sourcefunc}
Here we define the coefficients that we will use to write the absorption and emission coefficients $\alpha_\nu$ and $j_\nu$~\citep{rybickilightman}.
Assuming a power-law dependence of magnetic field and number density (equation~\eqref{eq:modelB},~\ref{eq:ne}), we have
\begin{align}
    \alpha_\nu&=\alpha_{\nu0}\nu^{-(p+4)/2}\left(\frac{z}{z_0}\right)^{-k_\alpha}\label{eq:alphanu}\\
    j_\nu&=j_{\nu0}\nu^{-(p-1)/2}\left(\frac{z}{z_0}\right)^{-k_j}\label{eq:jnu}
\end{align}
where
\begin{align}
    k_\alpha&=\frac b2(p+2)+a+d\to \frac12(p+6) \label{eq:knu}\\
    k_j&=\frac b2(p+1)+a+d\to\frac12(p+5)
\end{align}
Here, the arrow indicates the value for the canonical model parameters of $b=1$ and $a=2$.
The coefficients $\alpha_{\nu 0}$ and $j_{\nu0}$ depend on the power-law slope $p$ of the electrons and the values at the outflow base:
\begin{align}
    \alpha_{\nu0}(p)&=A_1f_1(p)f_4(p)B_0^{(p+2)/2}\eta_0 n_{e0}\\
    &=A_1f_1(p)f_4(p)\left(4\pi m_ec^2\right)^{(p+2)/4}\eta_0\sigma_{e0}^{(p+2)/4)}n_0^{(p+6)/4}\\
    j_{\nu0}(p)&=A_2f_2(p)f_4(p)B_0^{(p+1)/2}\eta_0 n_{e0}\\
    &=A_2f_2(p)f_4(p)\left(4\pi m_ec^2\right)^{(p+1)/4}\sigma_{e0}^{(p+1)/4}n_{e0}^{(p+5)/4}.
\end{align}
The constant $A_1$, $A_2$, and $p$-dependent quantities $f_1(p)$, $f_2(p)$, $f_4(p)$ are given by: 
\begin{align}
    A_1&=\frac{\sqrt{3}e^3}{8\pi m_e^2c^2} \label{eq:A1}\\ A_2&=\frac{\sqrt{3}e^3}{4\pi m_e c^2}\\
    f_1(p)&=\left(\frac{3e}{2\pi m_ec}\right)^{p/2}\Gamma\left(\frac{3p+2}{12}\right)\Gamma\left(\frac{3p+22}{12}\right) \\
    f_2(p)&=\frac{1}{p+1}\left(\frac{3e}{2\pi m_e c}\right)^{(p-1)/2}\Gamma\left(\frac p4+\frac{19}{12}\right)\Gamma\left(\frac p4-\frac{1}{12}\right)
\end{align}
where $e$ is the charge of an electron and $\Gamma(x)$ is the Gamma function.

In the optically-thick limit, the source function can be written as $S_\nu=j_\nu/\alpha_\nu$.
Combining Eqs.~\ref{eq:alphanu} and~\ref{eq:jnu} gives:
\begin{equation}
    S_\nu=S_{\nu0}\nu^{5/2}\left(\frac{z}{z_0}\right)^{-k_S}, \label{eq:snu}
\end{equation}
where
\begin{align}
    S_{\nu0}&=A_3f_3(p)R_\Gamma B_0^{-1/2}=A_3f_3R_\Gamma\left(4\pi m_ec^2\right)^{-1/4}\sigma_{e0}^{-1/4}n_{e0}^{-1/4}\label{eq:snu0}\\
    k_S&=-b/2\to-1/2.
\end{align}
The coefficients are given by:
\begin{align}
    A_3&=\left(\frac{8\pi m_e^3c}{3e}\right)^{1/2}\\
    f_3(p)&=\frac{1}{p+1}R_\Gamma\\
    R_\Gamma&=\frac{\Gamma(p/4+19/12)\Gamma(p/4-1/12)}{\Gamma(p/4+1/6)\Gamma(p/4+11/6)}.
\end{align}
Notably, the p-dependence in equation~\eqref{eq:snu} exists entirely in the $\nu$-independent function $f_3(p)$.
The normalization $C$ drops out of $S_\nu$ entirely.

\subsection{Side-on outflow} \label{app:sideon}
In this appendix, we provide details for the calculations in Sec.~\ref{ssec:rtoutflow}.
We assume that the outflow and its central black hole are located a distance $D$ away from the observer and that the emission and absorption coefficients are zero within this region.
We assume that $D\gg H_w\gg R_w$ and approximate all light rays as parallel to the observation plane.
See Figure~\ref{fig:schematic}.
For this edge-on case, the radiative transfer equation for the specific intensity $I_\nu$ coming from each outflow height is easily solvable because the emissivity and absorption coefficients are constant along the path length, e.g. depend only on $z$.
Under these assumptions,
\begin{equation}
    I_\nu(z)=S_\nu\left(1-e^{-\tau_\nu}\right),\label{eq:Inu}
\end{equation}
where the source function $S_\nu=j_\nu/\alpha_\nu$.
Synchrotron self-absorption dominates the absorption coefficient $\alpha_\nu$ and we assume that the emissivity $j_\nu$ comes from the nonthermal electrons' synchrotron emission~\citep[equation 6.36, 6.53]{rybickilightman}, yielding (equation~\eqref{eq:snu0}):
\begin{equation}
    S_\nu(z)\sim \nu^{5/2}B_\perp^{-1/2}.\label{eq:snuscale}
\end{equation}

The optical depth $\tau_\nu$ at a given height and frequency $\nu$ is:
\begin{equation}
  \tau_\nu(z)\equiv\int_{-R_w(z)}^{R_w(z)}\alpha_\nu(s)~{\rm d}s=2R_w(z)\alpha_\nu(z).\label{eq:taunu}
\end{equation}

The total flux from the outflow is given by the sum over outflow heights:
\begin{align}
  F_\nu&=\int I_\nu\sin i~{\rm d}\Omega=\int I_\nu~{\rm d}\Omega,\label{eq:Fnu}\\
  &=\frac{1}{ D^2}\int_{z_0}^{H_w+z_0}2R_w(z')I_\nu(z')~{\rm d} z'\\
  &=\frac{2}{ D^2}\int_{z_0}^{H_w+z_0}R_w(z')S_\nu(z')\left(1-e^{-2R_w(z')\alpha_\nu(z')}\right)~{\rm d}z'
\end{align}
where
\begin{equation}
  {\rm d}\Omega=\frac{1}{D^2}\int_{-R_w}^{R_w}{\rm d}y\int_{z_0}^{H_w+z_0}{\rm d}z=\frac{1}{ D^2}\int_{z_0}^{H_w+z_0}2R_w(z){\rm d}z \label{eq:dOmega}
\end{equation}
because the emission properties are independent of the third dimension $y$.

\subsection{Face-down outflow} \label{app:facedown}

\begin{figure}
    \centering
    \includegraphics[width=0.45\textwidth]{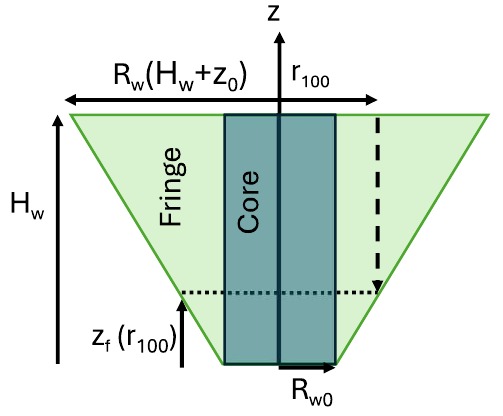}
    \caption{Diagram of the face-down outflow set-up showing the presence of the outflow ``core'' and ``fringes'' that form the outflow's flat spectrum. The coordinate $r$ runs perpendicular to the jet axis and to the face-down observer rays. The outflow core runs from $r=0$ to $r=R_{w0}$ and has the same turnover frequency for this range of $r$. The outflow fringe turnover frequency decreases with $r>R_{w0}$ due to the smaller path length through the outflow, with rays extending from the outflow top down to $z_f(r)$ (equation~\eqref{eq:rc}). The diagram shows the outflow radius $r_{100}$, the projected size of the 100 GHz emission for a face-down observer. For clarity, the black hole and accretion disc have been removed, though see Figure~\ref{fig:schematic}. Not to scale.}
    \label{fig:facedown_diagram}
\end{figure}

In this appendix, we characterize the limiting case where the observers views the outflow face-down.
With this viewing angle, the source function is no longer constant with height.
If the outflow width were independent of height, the spectrum would no longer feature the intermediate regime between the purely optically-thin and purely optically-thick regimes where the heights stack together to produce a flatter spectrum.
However, because the outflow's radius increases with $z$, there is a ``core'' whose rays travel all the way back to the outflow base $z_0$ as well as ``fringes'' where only the upper parts of the outflow contribute; see Figure~\ref{fig:facedown_diagram}.
Each fringe will have the two-component optically-thick and optically-thin spectrum, and will sum with fringes at different $x$ and $y$ coordinates.
The face-down scenario therefore also features the intermediate spectrum where the spectral slope lies between the optically-thin and optically-thick limits. 

To calculate the contribution of the outflow fringes to the total flux, we treat the outflow as spherical rather than square; the difference in flux will be order unity.
We first define the outflow's cylindrical radius $r\equiv\sqrt{x^2+y^2}$, which runs from $R_{w0}$ to $r_{\rm max}(z)=R_{w0}(z/z_0)^w$.
Then, for a given position $r$ at the outflow top, a given ray will run through the outflow down to a height $z_f$, where
\begin{equation}
    z_f(r)=z_0\left(\frac{r}{R_{w0}}\right)^{1/w}. \label{eq:rc}
\end{equation}
To find $I_\nu$ and therefore the turnover frequency of the outflow fringes, we calculate the optical depth:
\begin{equation}
    \tau_\nu(r)=\int_{z_f(r)}^{H_w+z_0}\alpha_\nu(z')dz'=\alpha_{\nu0}\nu_t^{-(p+4)/2}\int_{z_f(r)}^{H_w+z_0}\left(\frac{z'}{z_0}\right)^{-k_\alpha}dz'.
\end{equation}
Setting $\tau_\nu(\nu_t)=1$, we find
\begin{align}
    \frac{\nu_t(r)}{\nu_t(r<R_{w0})}&=\left(\frac{r}{R_{w0}}\right)^{\frac{2(1-k_\alpha)}{w(p+4)}}\to \left(\frac{r}{R_{w0}}\right)^{-1/w}\label{eq:facedownrc}\\
\end{align}
where $\nu_t(r<R_{w0})=\alpha_{\nu0}z_0/(k_\alpha-1)$, with $k_\alpha$ given by equation~\eqref{eq:knu}.
Note that for $a=2$ and $b=1$, the drop-off of the turnover frequency with height is independent of the electron distribution function's power-law index $p$. 

\begin{figure*}
    \centering
    \includegraphics[width=0.95\textwidth]{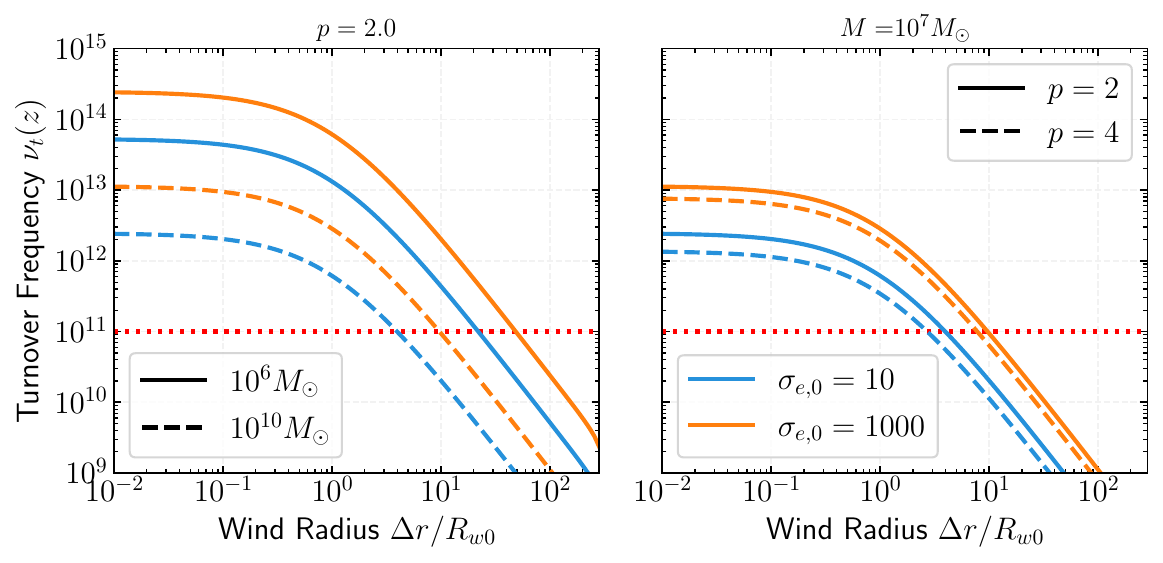}
    \caption{For a face-down outflow, the turnover frequency of a fringe with distance $\Delta r\equiv r-R_{w0}$ from the core (equation~\eqref{eq:facedownrc}) varies with height. The left and right panels show dependence of $\nu_t(z)$ on the black hole mass and electron power-law index, respectively. In both panels, colors show different values of the electron magnetization at the outflow base $\sigma_{e0}$. See e.g. equation~\eqref{eq:nut_facedown}. Horizontal dotted red lines show where the outflow emits $100~{\rm GHz}$ radiation. 
    }
    \label{fig:nut_facedown}
\end{figure*}

The value of the core's turnover frequency is:
\begin{align}
    \nu_t(r<R_{w0}) & \stackrel{\mathclap{\mbox{p=2}}}{=}5.2\times10^{13}~{\rm Hz}\left(\frac{\sigma_{e0}/100}{M/10^7M_\odot}\right)^{1/3}\\
    &\stackrel{\mathclap{\mbox{p=3}}}{=}3.4\times10^{13}~{\rm Hz}\left(\frac{\sigma_{e0}/100}{M/10^7M_\odot}\right)^{5/14} \label{eq:nut_facedown}
\end{align}
These values are quite similar to the equivalent values for a side-on outflow (equation~\eqref{eq:nut_sigmaM_scaling}) because although each ray traverses the entire outflow, the absorption is dominated by the base where the magnetic field is strongest.
Figure~\ref{fig:nut_facedown} shows how the distance from the outflow's core changes the turnover frequency.

In analogy to the side-on outflow's height where $\nu_t=100~{\rm GHz}$, ($z_{100}$; equation~\eqref{eq:z100}), we define the radius $r_{100}$ such that $\nu_t(r_{100})=100~{\rm GHz}$. 
This radius will be the projected size of the mm emission for a face-down observer.

The total flux from the outflow is given by the sum over the outflow's face:
\begin{align}
  F_\nu(r)&=\frac{1}{4\pi D^2}\int I_\nu r'~{\rm d}r'~{\rm d}\phi=\frac{2\pi}{4\pi D^2}\int_0^{r}I_\nu(r')~{\rm d}r',\label{eq:FnuFaceDown}\\
  &=\frac{2\pi}{D^2}\left[\int_0^{R_{w0}}I_\nu^{\rm core}r'~{\rm d}r' + \int_{R_{w0}}^r I_\nu^{\rm fringe}(r')r'~{\rm d}r'  \right]\\
  &=\frac{1}{2D^2}\left[\frac{R_{w0}^2}2I_\nu^{\rm core}+ \int_{R_{w0}}^r I_\nu^{\rm fringe}(r')r'~{\rm d}r'  \right]
\end{align}
where the specific intensity $I_\nu$ is given by solving the radiative transfer equation:
\begin{align}
    I_\nu(r, z)&=e^{-\int_{z_f(r)}^z \alpha_\nu(z')~{\rm d}z'}\int_{z_f(r)}^z j_\nu(z') e^{\int_{z_f(r)}^{z'}\alpha_\nu(z'')~{\rm d}z''}~{\rm d}z'\\
    &=e^{-\tau_\nu(r, z)}\int_{z_f(r)}^z j_\nu(z')e^{\tau_\nu(r,z)}~\rm {d}z'.
\end{align}

We can estimate the total flux's dependency on $\nu$ by using $\nu_c\sim r_c^{-1/w}$ (equation~\eqref{eq:facedownrc}) and $z_c\sim r_c^{1/w}\sim \nu_c^{-1}$ (equation~\eqref{eq:rc}).
Thus we have approximately:
\begin{align}
    F_\nu&\propto {\rm area}\times S_\nu\propto r_c^2\nu_c^{5/2}z_c^{1/2}\propto (\nu_c^{-w})^2\nu_c^{5/2}\nu_c^{-1/2}\\
    &\propto \nu_c^{2(1-w)}.\label{eq:facedownalpha}
\end{align}
Notably, the spectral index is always independent of $p$ since $\nu_t(z)$ is (equation~\eqref{eq:facedownrc}). 
For $w=1$, the flux is flat with frequency, same as for the side-on outflow (equation~\eqref{eq:alphaF}).
For a parabolic outflow with $w=1/2$, $F_\nu\propto\nu$, i.e. $\alpha=-1$.


\bibliography{refs}{}
\bibliographystyle{aasjournalv7}

\end{document}